\documentclass[sn-basic]{sn-jnl}

\usepackage{graphicx}%
\usepackage{multirow}%
\usepackage{amsmath,amssymb,amsfonts}%
\usepackage{amsthm}%
\usepackage{mathrsfs}%
\usepackage[title]{appendix}%
\usepackage{xcolor}%
\usepackage{textcomp}%
\usepackage{manyfoot}%
\usepackage{booktabs}%
\usepackage{algorithm}%
\usepackage{algorithmicx}%
\usepackage{algpseudocode}%
\usepackage{listings}%

\makeatletter
\input{aas_macros.sty}
\def\ref@jnl#1{{\jnl@style#1\ }}
\makeatother

\begin{document}

\title[Active Galactic Nuclei as high-energy neutrino sources]{Active Galactic Nuclei as high-energy neutrino sources}


\author*[1]{\fnm{Filippo} \sur{D'Ammando}}\email{dammando@ira.inaf.it}



\affil*[1]{\orgdiv{IRA Bologna}, \orgname{INAF}, \orgaddress{\street{Via gobetti 101}, \city{Bologna}, \postcode{40129}, \state{Bologna}, \country{Italy}}}




\abstract{Identifying the sources of the high-energy astrophysical neutrinos has been one of the main topics in astrophysics since the first observation of high-energy neutrinos by the IceCube Neutrino Observatory. Active Galactic Nuclei (AGN) are sources of high-energy $\gamma$-rays and are considered to be promising candidates to be sources of high-energy neutrinos and ultra-high energy cosmic rays as well. However, several studies suggest that the neutrino emission from the $\gamma$-ray blazar population only accounts for a small fraction of the total astrophysical neutrino flux. We present and discuss recent results on the search for correlations between astrophysical neutrinos and both $\gamma$-ray and radio bright AGN. 
The IceCube Collaboration has reported high-energy neutrino events that may come from both the radio-loud AGN TXS\,0506+056 and the radio-quiet AGN NGC\,1068. Other cases of possible associations between high-energy neutrino events and individual blazars were claimed with controversial results. We discuss the properties of these sources together with the different neutrino production mechanisms proposed for those sources. Finally, we outline future prospects in the field, focusing on remaining open questions, the development of upcoming neutrino facilities, and the evolving multi-frequency landscape within the multi-messenger era.}


\maketitle

\clearpage
\setcounter{tocdepth}{3} 
\tableofcontents

\section{Introduction}\label{sec1}

Alongside gravitational waves, high-energy neutrinos have opened up a new chapter in astronomy. Except for photons, neutrinos are the most abundant particles in the Universe, but because they do not have an electrical charge and almost no mass, they are elusive particles. In particular, thanks to their small interaction cross sections, they can escape dense astrophysical sources that may be opaque to photons with paths unaffected by intervening electromagnetic fields. 

The first detection of MeV neutrinos from a non-solar astrophysical source was from Supernova 1987A (SN1987A) in the Large Magellanic Cloud on 1987 February 23 by the Kamiokande and Irvine-Michigan-Brookhaven detectors, revealing core-collapse processes and pioneering neutrino astronomy \citep{scholberg12, tamborra18}, while solar neutrinos were first detected earlier by Raymond Davis Jr.'s experiment, confirming solar fusion, with later detectors like Borexino fully mapping the Sun's reactions \citep{haxton13}, bridging solar and supernova neutrino observations into a new era of cosmic study. 

Over the past decade, neutrino astronomy has emerged as a new and exciting window for exploring the most energetic phenomena in the Universe. The current-generation of facilities started to detect high-energy neutrinos of astrophysical origin, opening the field to identification of objects that produce such neutrinos and thus to new exciting discoveries. The confirmation of high-energy cosmic neutrinos validates the hypothesis that astrophysical sources can accelerate particles to incredibly high energies, opening the field to transformative discoveries.

A first milestone occurred in 2013, with the detection of a diffuse neutrino flux by the IceCube Neutrino Observatory (hereafter IceCube) \citep{aartsen13}, which started a new era of neutrino astronomy. This is part of a larger revolution in the field, and the association between a high-energy neutrino and the blazar TXS\,0506+056, being the first time an extragalactic neutrino has been traced to its source, represented a further fundamental step. In 2022, IceCube identified NGC\,1068 as the first steady-state extragalactic source of high-energy neutrinos \citep{abbasi22b}. NGC\,1068 is a ``non-jetted'' or ``radio-quiet'' Active Galactic Nucleus (AGN), indicating that different types of AGN, not just blazars, are high-energy cosmic accelerators. In 2025, Neutrino astronomy has garnered even greater attention due to the announcement by the KM3NeT Collaboration of the observation of an ultra-high-energy neutrino, with an energy estimated of $\sim$ 220 PeV, which marks the highest energy neutrino ever detected \citep{aiello25}. These detections represent a pivotal moment, opening a new channel for multi-messenger astronomy and providing information that complements observations made with electromagnetic signals like photons and cosmic rays. However, pinpointing the precise sources of these cosmic neutrinos remains a significant challenge.  

Neutrinos at TeV--PeV energies are typically produced when relativistic protons interact with matter via inelastic $p$--$p$ scattering or with radiation via photohadronic processes, $p$--$\gamma$ interactions. Such processes also produce neutral pions that decay into $\gamma$-rays. The simultaneous detection of neutrinos and $\gamma$-rays can provide fundamental insight into the nature of production sources. Current observations reveal a widespread, diffuse flux of neutrinos, suggesting contributions from a variety of sources.

AGN are considered the most luminous persistent sources of high-energy radiation and promising candidate neutrino sources \citep[e.g.,][]{stecker91, mannheim92, szabo94, atoyan01}. AGN are powered by a super-massive black hole (SMBH) actively consuming matter, forming a hot, luminous accretion disc and a surrounding corona, all enveloped by a donut-shaped dusty torus of gas and dust. In about 10\% of AGN, usually termed radio-loud AGN, the accretion disc is the base of a bipolar outflow of relativistic plasma, which may extend well beyond the host galaxy, forming the spectacular lobes visible in the radio band.
In particular, AGN with relativistic jets are the most powerful persistent astrophysical sources of electromagnetic radiation in the Universe, therefore promising sites of cosmic-ray acceleration and thus neutrinos in interactions with photon fields or matter. In this context, blazars are the most extreme subclass of AGN with jets closed aligned to our line of sight. Their emission dominates the extragalactic $\gamma$-ray sky reaching energies up to TeV, accelerating particles to very high energies. Due to the small viewing angle ($< 10^{\circ}$) the electromagnetic emission and potential neutrino emission from the jet become strongly boosted, making blazars among the brightest sources of photons in the Universe and thus prime targets for a population-based neutrino counterpart search. Blazars are characterized by a spectral non-thermal continuum from radio to $\gamma$-rays, high polarization degree, apparent super-luminal motion of jet components observed in radio images, and rapid variability at all wavelengths down to a few minutes. Blazars are usually divided into two classes, BL Lac objects (BL Lac) and Flat Spectrum Radio Quasars (FSRQ), based on the presence of weak or strong emission and absorption lines in their optical and UV spectra \citep[e.g.,][]{stickel91}. This is related to a different accretion regime \citep{ghisellini11}, which can have a different impact on the inflow and outflow of the source. This is similar to the dichotomy observed in the proposed parent population of radio galaxies, where High-Excitation Radio Galaxies (HERG) operate via radiatively efficient, geometrically thin accretion discs fueled by cold gas, whereas Low-Excitation Radio Galaxies (LERG)  operate via radiatively inefficient, geometrically thick flows \citep[e.g.,][]{best12}. In this context, LERGs are primarily linked to faint jets in Fanaroff--Riley~I, associated with BL Lacs, and HERGs are predominantly linked to powerful jets in Fanaroff--Riley~II, associated with FSRQs \citep[e.g.,][]{fanaroff74}. 

Several multi-wavelength monitoring campaigns have recently been triggered in response to high-energy neutrinos observed with the IceCube Neutrino Observatory from the direction of blazars.
In the first years of IceCube operation, any systematic search for the temporal and spatial association of AGN with neutrino clustering has not revealed any significant detection \citep[e.g.,][]{aartsen17, aartsen20}, therefore a search for variable sources over the electromagnetic spectrum temporally and spatially coincident with high-energy neutrinos detected by IceCube has been performed. So far, an association between high-energy neutrinos and a source has been confirmed at $>$ 3$\sigma$ level only for the blazar TXS\,0506+056 and the Seyfert II galaxy NGC\,1068 \citep{abbasi23c}, in addition to the Galactic plane \citep{abbasi23c}. The detection of high-energy neutrinos from NGC\,1068 by IceCube opened new theoretical scenarios by pointing to AGN as cosmic neutrino factories, particularly suggesting production in the dense corona surrounding its SMBH, challenging purely leptonic models and strongly favoring hadronic processes involving proton-photon interactions in high magnetic fields.

In this review, we report recent results from studies of high-energy astrophysical neutrinos and their potential extragalactic counterparts, specifically AGN, from an observational and phenomenological point of view. In Sect.~\ref{sec1b}, we briefly describe the emission mechanisms producing high-energy neutrinos in AGN. More details about theoretical studies of the processes at work in neutrino emitting AGN can be found e.g. in \citet{halzen02, bottcher13, cerruti20, murase23}. In Sect.~\ref{sec2}, we describe the properties of the confirmed neutrino-associated AGN, TXS\,0506+056 and NGC\,1068, followed by a discussion of other potential associations in Sect.~\ref{sec3}. Section~\ref{sec4} focuses on results from searches for multi-flare neutrino emission from AGN. Sections \ref{gamma} and \ref{radio} describe the correlations between neutrinos and $\gamma$-ray-emitting and radio-bright AGN, respectively. In Sect.~\ref{sec7}, we discuss future prospects, including open questions, neutrino facilities, and multi-frequency observations. Finally, Sect.~\ref{sec8} provides concluding remarks.

\section{Emission mechanisms producing high-energy neutrino in AGN}\label{sec1b}

The spectral energy distribution (SED) of blazars is typically double-humped with a first peak that is due to synchrotron radiation from high energy leptons (electrons/positron) in a relativistic jet, while the origin of the second peak is still debated \citep[e.g.,][]{ghisellini09, ghisellini17}. In the leptonic scenario, it has been interpreted as inverse Compton (IC) scattering off seed photons, either internal or external to the jet, by highly relativistic electrons \citep[e.g.,][]{ulrich97}, although other models involving hadronic processes or lepto-hadronic processes have gained more and more ground \citep[e.g.,][for a review]{bottcher13,cerruti15,petropoulou15}. 

In the hadronic scenario, the processes are associated with protons and higher-Z nuclei. Protons can radiate directly via synchrotron emission or indirectly via secondary particles produced by photon-proton interactions. In the second case, two important mechanisms are photo-meson production and Bethe-Heitler pair production. The photo-meson (or photo-pion) mechanism refers to the production of pions, in particular neutral and charged pions, through the interaction of high-energy photons with matter (typically nucleons or nuclei). Neutral pions then decay directly into photons, while charged pions decay into leptons and neutrinos. Additionally, photons and secondary leptons from pion decay can produce pair cascades that can emerge as a secondary radiative component in the SED. Furthermore, the Bethe-Heitler process is the direct production of an electron/positron pair and triggers a pair cascade in the emitting region. The Bethe-Heitler process can be the main proton-photon interaction process in the case of low-energy photons, while increasing the proton energy the photo-meson process becomes dominant. In principle, neutrinos can also be produced in proton-proton interactions, but, due to the low density of plasma in the AGN jets, this process is negligible in AGN (except in the case of target photons coming from broad line region (BLR) clouds or stars interacting with the jet). In case of proton-synchrotron radiation, with respect to lepton-synchrotron radiation, the magnetic field, the particle density, or both have to be significantly increased to achieve comparable radiation efficiency. Another general problem with the hadronic models, in particular if one wants to maximize the neutrino output, is the power associated with the hadronic emission. This energetic issue is usually more significant for FSRQ, where proton kinetic powers can reach an unphysical, super-Eddington regime \citep[e.g.,][]{bottcher13, zdziarski15}. 

From an observational point of view, the synchrotron peak occurs in the IR/optical band in FSRQ and at UV/X-rays in BL Lacs, whereas the high-energy peak occurs in the MeV-GeV regime in FSRQ and in the GeV-TeV regime in BL Lacs. Different target photons are important in the multi-messenger context for BL Lacs and FSRQ: in the first case synchrotron photons from relativistic electron accelerated, in the second one external radiation fields (e.g. broad line emission, dust emission, disc-corona emission) are important for producing neutrino emission. Mechanisms of energy dissipation and particle acceleration in the jet are still under debate, with internal shocks and shock acceleration can be historically considered, and magnetic reconnection and stochastic acceleration recently considered as the main mechanism of particle acceleration. In case of shock acceleration and magnetic reconnection not only electrons but also ions are accelerated. A detailed discussion of particle acceleration in AGN jets is out of the scope of this review, see e.g. \citet{kirk00, sironi15, comisso18, blandford19}.

The main emission mechanisms for producing high-energy neutrinos in non-jetted AGN, such as Seyfert galaxies, are hadronic processes, specifically proton-proton interactions and photo-hadronic interactions \citep[see e.g.][]{murase20b, murase22, das24}. These interactions occur as cosmic rays (protons) are accelerated in dense, extreme environments near the central SMBH and collide with ambient gas or radiation fields. The most promising sites of neutrino production are the magnetized corona, fast AGN-driven winds, and to a lesser extent failed winds. The hot, magnetized corona above the accretion disc can accelerate protons via magnetic reconnection and/or turbulence. These protons interact with intense X-ray, UV, and optical radiation fields from the disc (photo-meson or proton-photon interactions) or with the plasma itself to produce neutrinos. Alternatively, fast AGN-driven winds or outflows can collide with dense, clumpy gas clouds in the circumnuclear region, generating bow shocks that accelerate protons. These accelerated protons then interact with the colder, ambient protons of the outflowing and surrounding gas to produce pions, which decay into neutrinos. A secondary mechanism involves acceleration in "failed" winds, where material falls back or stalls, generating collisionless shocks and dense, turbulent regions and providing potential target gas and photon density to produce neutrinos \citep{inoue22}.

\section{Extragalactic sources neutrino emitters}\label{sec2}

\subsection{TXS\,0506+056: the first association of a high-energy neutrino with an AGN}

\subsubsection{IceCube-170922A}

\begin{figure}[ht]
\centering
\includegraphics[width=\textwidth]{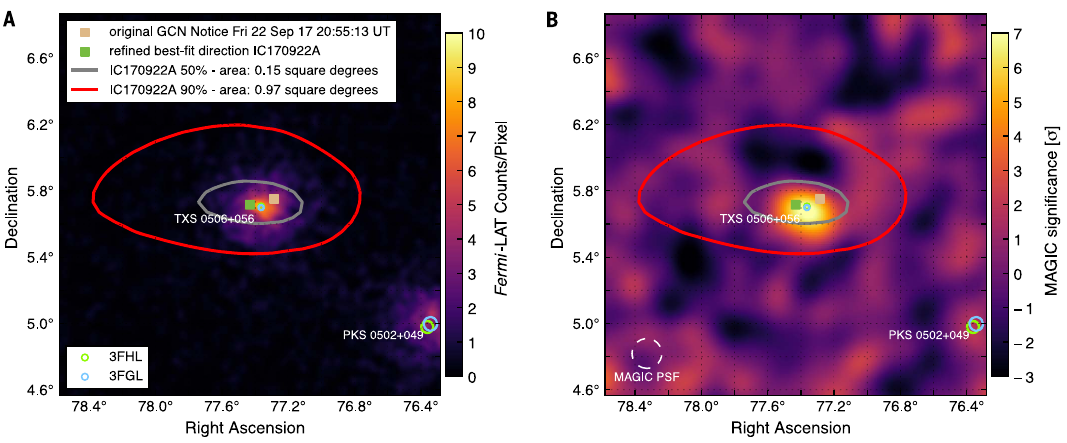}
\caption{{\it Left:} Sky position of the neutrino IceCube-170922A in equatorial coordinates overlapped with the counterpart observed by \textit{Fermi}-LAT (left panel) and MAGIC (right panel) telescopes. The tan square indicates the position reported in the initial alert, and the green square indicates the final best-fitting position from follow-up reconstructions. Gray and red curves show the 50\% and 90\% neutrino containment regions, respectively, including statistical and systematic errors. Image reproduced with permission from \citet{aartsen18c}, copyright by AAAS}\label{0506_fig1}
\end{figure}

On 2017 September 22, the IceCube neutrino observatory detected a 290 TeV neutrino (IceCube-170922A) from a direction consistent with the blazar TXS\,0506+056 (with a positional uncertainty of 0.4--0.8$^{\circ}$) found to be in a $\gamma$-ray flaring state both by the Large Area Telescope (LAT) on board the \textit{Fermi Gamma-ray Space Telescope} and the Major Atmospheric Gamma Imaging Cherenkov Telescopes (MAGIC) \citep{aartsen18c} (see Fig.~\ref{0506_fig1}). In particular, \textit{Fermi}-LAT detected the source in a six month-long flaring state, while MAGIC detected a fast flare at Very High Energies (VHE) about 10 days after the IceCube neutrino detection. The chance probability that the IceCube event and the $\gamma$-ray flaring activity occurred simultaneously in space and time was estimated at the level of 3-$\sigma$. This multi-messenger photon-neutrino emission has been the smoking gun attesting to the presence of highly relativistic hadrons in AGN jets, and currently it has emerged as the second-most prominent hotspot in the neutrino sky over 10 years of observations.

\begin{figure}[ht]
\centering
\includegraphics[width=\textwidth]{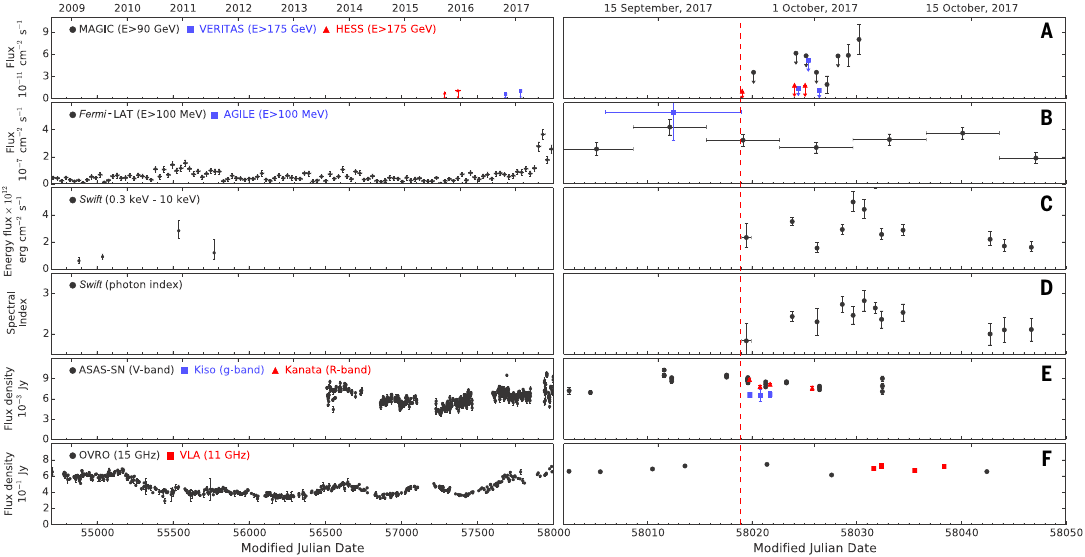}
\caption{Multi-wavelength observations of TXS\,0506+056 before and after IceCube-170922A including from top to bottom VHE observations by MAGIC, H.E.S.S., and VERITAS; HE observations by \textit{Fermi}-LAT and AGILE; X-ray observations by \textit{Swift}-XRT; optical observations by ASAS-SN, Kiso/KWFC, and Kanata/HONIR; and radio observations by OVRO and VLA. The red dashed line marks the detection time of the neutrino IceCube-170922A. Image reproduced with permission from \citet{aartsen18c}, copyright by AAAS}\label{0506_fig2}
\end{figure}

Before 2017, TXS\,0506+056 was little monitored in the electromagnetic spectrum and then studied. The neutrino association triggered several multi-frequency observations, covering the entire electromagnetic spectrum (see Fig.~\ref{0506_fig2}), and detailed studies of the source. The redshift of the source has been tentatively estimated as $z = 0.3365 \pm 0.0010$, based on three very weak emission lines observed with the Gran Telescopio Canarias identified with [OII]3727 \AA, [OIII]5007 \AA, and [NII]6583 \AA\, \citep{paiano18}. In the radio band, the Owens Valley Radio Observatory (OVRO) has been monitoring the source for several years and observed an increase in the flux density in early 2017. The source has also been monitored at 15 GHz with Very Long Baseline Interferometry (VLBI) observations within the Monitoring of Jets in Active Galactic Nuclei with VLBA Experiments \citep[MOJAVE;][]{lister09}, which reports a maximum apparent jet speeds of $(0.98 \pm 0.31)c$ \citep{lister19}. Such low jet velocities are frequently reported for a number of TeV emitting BL Lac objects \citep[][and the reference therein]{piner14, piner18}.

\begin{figure}[ht]
\centering
\includegraphics[width=0.5\textwidth]{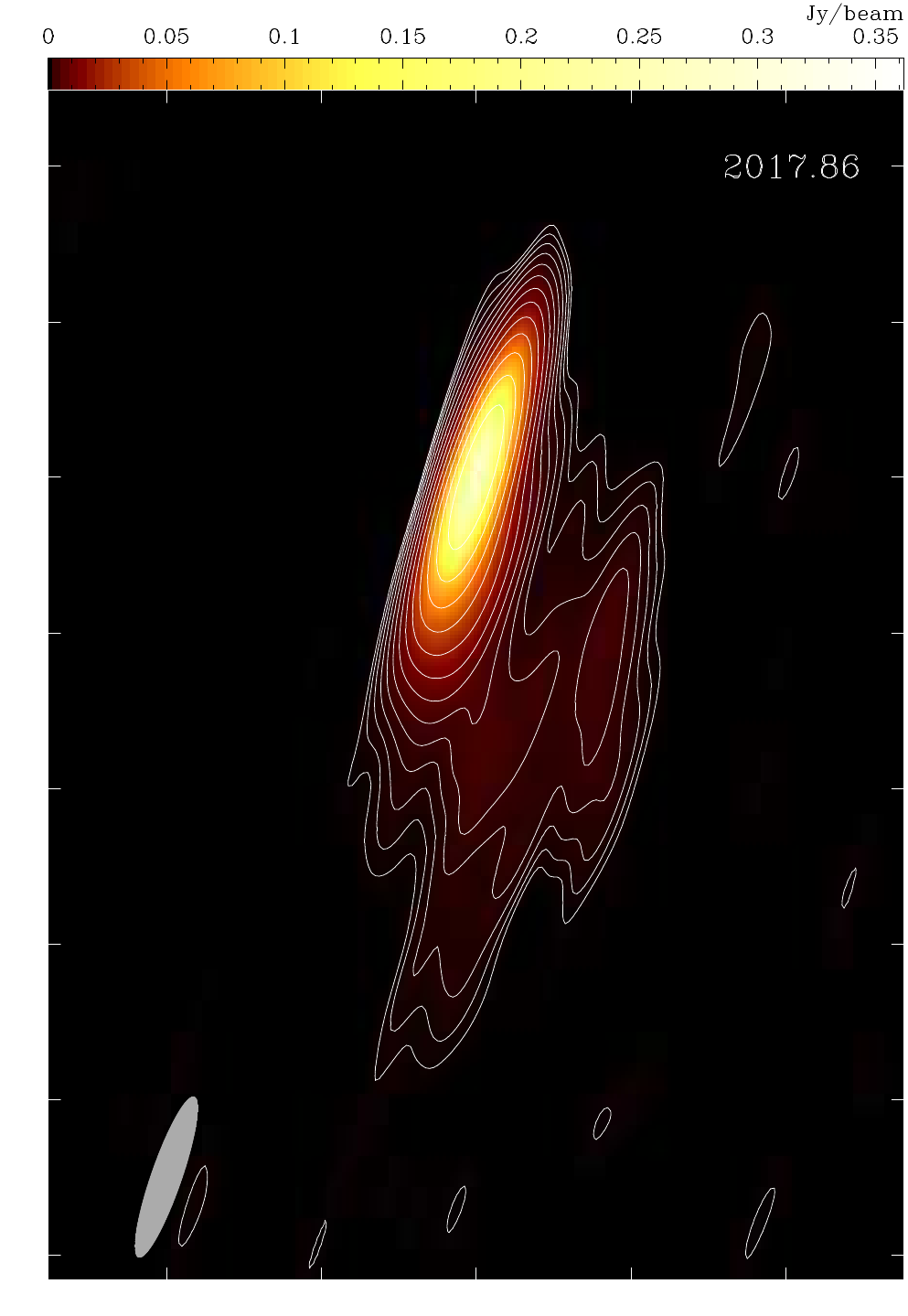}\hfill
\includegraphics[width=0.5\textwidth]{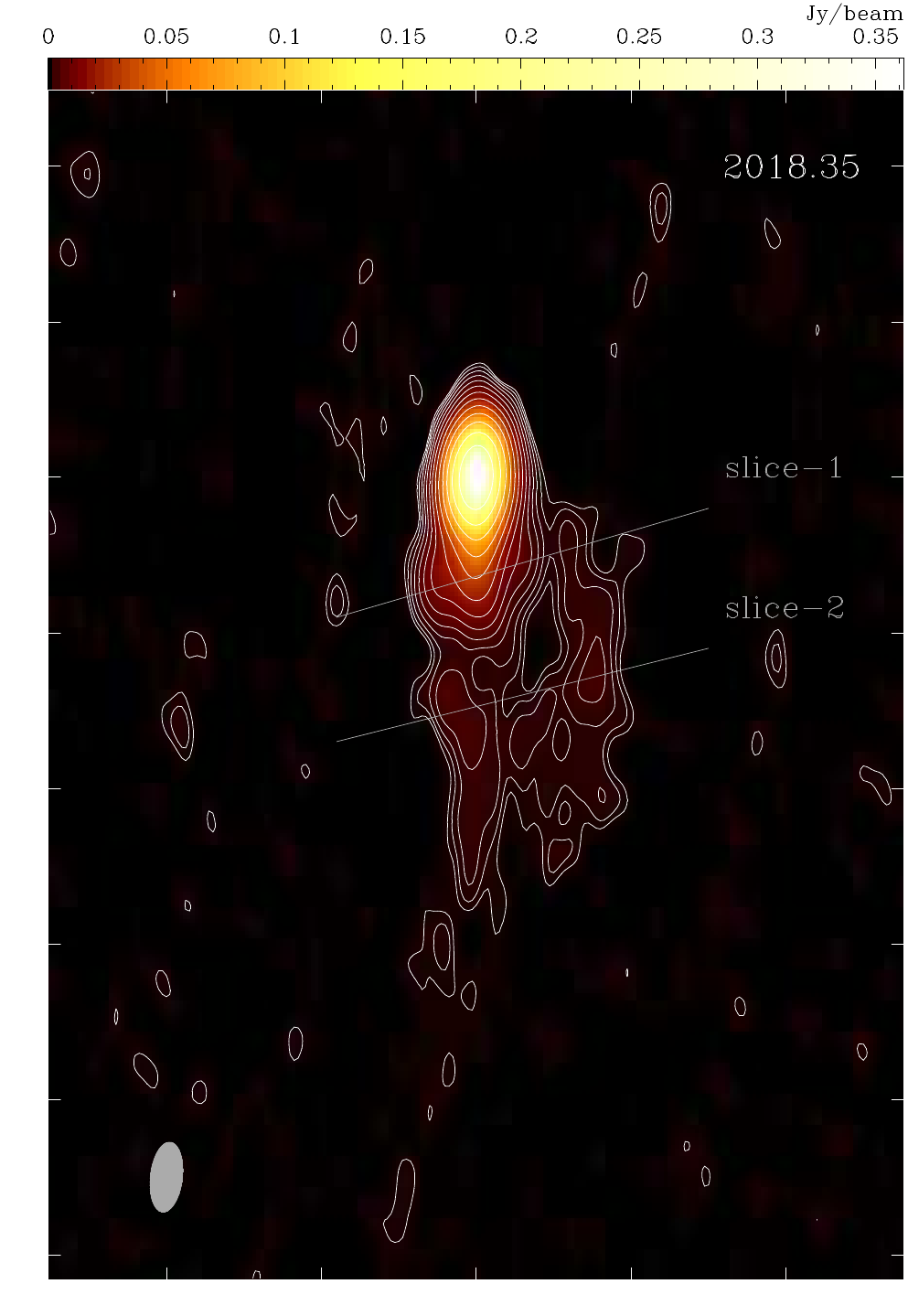}
\caption{43 GHz VLBA images of TXS\,0506+056 collected on 2017 November 11 {\it (left panel)}  and 2018 May 4 {\it (right panel)}. Image reproduced with permission from \citet{ros20}, copyright by the author(s)}\label{0506_fig6}
\end{figure}

Based on the radio and OII luminosities, the emission line ratios and the Eddington ratio, this source has been proposed as a masquerading BL Lac, i.e. intrinsically a FSRQ with hidden broad lines and a standard accretion disc \citep{padovani19}. In addition, \citet{padovani19} proposed it as an atypical high-luminosity/high-synchrotron-peaked blazar, with a synchrotron peak located at 10$^{15}$ Hz, and a peak luminosity of $\sim$10$^{47}$ erg s$^{-1}$, a typical value for FSRQ. High-resolution Very Long Baseline Array (VLBA) radio images at 43 GHz show signs of an apparent limb brightening with a spine-layer structure within 1 mas from the mm-VLBI core of the source \citep{ros20} (Fig.~\ref{0506_fig6}). The slower flow is due to either general deceleration or jet-transversal velocity stratification \citep[e.g.,][]{tavecchio08, tavecchio15}. This spine-layer structure is consistent with theoretical models in which neutrino and $\gamma$-rays are produced by electrons and protons interactions in the highly relativistic jet spine with external photons originating from a slower moving jet layer. 

\subsubsection{Neutrino emission observed from TXS\,0506+056 during 2014/2015 without electromagnetic counterpart}

\begin{figure}[ht]
\centering
	\includegraphics[width=0.49\textwidth]{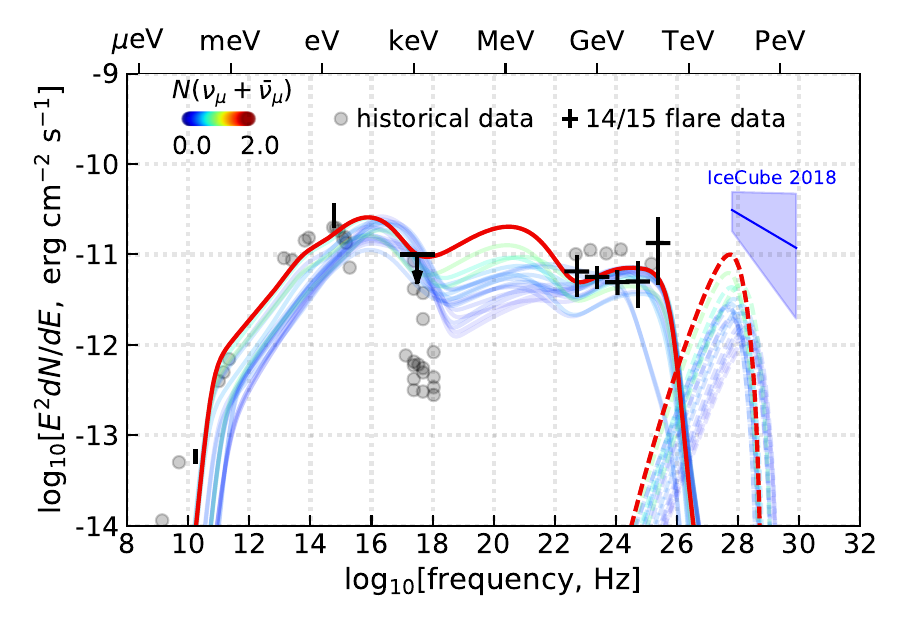}
	\includegraphics[width=0.49\textwidth]{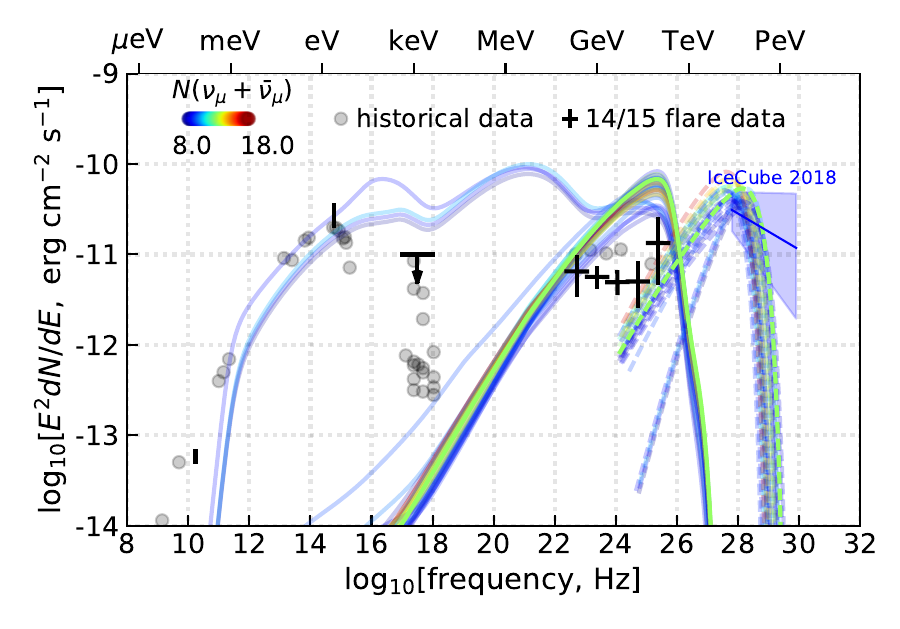}
\caption{SED and muon neutrino fluxes of TXS\,0506+056 during the 2014--2015 events predicted by the one-zone hadronic model, compared to the single-flavor flux derived by IceCube.  In the left panel, the parameters optimized to describe the SED in agreement with the electromagnetic observations fail to explain the neutrino emission, while in the right panel the parameter sets account for the muon neutrinos observed by IceCube, but overshoot the multi-wavelength emission. Historical data are reported in grey. Image reproduced with permission from \citet{rodrigues19}, copyright by AAS}\label{0506_fig5}
\end{figure}

In response to the 2017 association between the high-energy neutrino and TXS\,0506+056, an investigation has been performed over 9.5 years of IceCube neutrino observations to search for excess emission at the position of the blazar.
Analyzing IceCube data before the IceCube-170922A event, a neutrino flaring period of $\sim$110 days from TXS\,0506+056 was observed during 2014--2015 with a 3.7-$\sigma$ significance level, assuming a Gaussian time window \citep{aartsen18a}. In that case, no corresponding flaring activity has been detected in $\gamma$-rays \citep{garrappa19}. Moreover, the Burst Alert Telescope (BAT) on board the \textit{Neil Gehrels Swift Observatory (Swift)} was not triggered during its all-sky monitoring in the neutrino flare period and did not detect TXS\,0506+056 in the 15--50 keV band, implying a flux significantly less than 3 mCrab. Theoretical models with a single zone struggle to explain the 2014--2015 neutrino flare of TXS\,0506+056 due to constraints from X-ray data \citep{keivani18, murase18, rodrigues19, petropoulou20}. One potential model to explain both the 2014--2015 neutrino flare and the IceCube-170922A event is the emergence of a relativistic neutral beam in the blazar jet due to interactions of accelerated cosmic rays with photons \citep{zhang20}. Models invoking two radiation zones have been suggested to explain the 2014--2015 neutrino flare, in particular an IC-dominated compact-core model or a model involving an external radiation field from accretion disc radiation isotropized in BLR \citep[see Fig. \ref{0506_fig5};][]{rodrigues19}, although even in these models it is challenging to explain more than five neutrino events without violating the observational X-ray constraints or the observed $\gamma$-ray fluxes, respectively. It is worth noticing that the multi-wavelength coverage of the source during that period was very sparse, thus preventing a detailed characterization of the SED.

Other two interesting neutrino events from TXS\,0506+056 have been observed in 2021 and 2022. A cascade neutrino event from the direction of the source with an estimated energy of 224 $\pm$ 75 TeV was detected by the new Baikal-GVD neutrino telescope in 2021 April \citep[GVD-210418CA;][]{allakhverdyan24}. This event has a high probability to be of astrophysics origin, with a signalness parameter of 97.1\%, and has been tentatively associated with a radio flare. Moreover, this neutrino event has demonstrated the effectiveness in accurately identifying the direction of arrival of neutrinos by exploiting cascading events. Furthermore, IceCube reported a neutrino event with an estimate energy of $\sim$ 170 TeV in spatial coincidence with the source on 2022 September 8 \citep[IC-220918A;][]{becker22}. These neutrino events have provided independent confirmation of the association of TXS\,0506$+$056 with high-energy neutrinos.

In the end, the neutrino luminosity of TXS\,0506+056 is dominated by the 2014--2015 neutrino flare, with the IceCube 170922 event being a subdominant contribution. Finally, although an analysis of 10-yr IceCube observations of muon neutrinos has shown that TXS\,0506+056 is spatially coincident with the second brightest hotspot in the neutrino sky \citep{aartsen20}, suggesting that this blazar may also be a steady emitter of neutrinos up to PeV energies, the total contribution of the neutrino emission from TXS\,0506+056 to the diffuse astrophysical flux observed by IceCube is a few percent at most \citep{aartsen18a,aartsen18b}.

\subsubsection{Multi-wavelength monitoring of TXS\,0506+056 after 2017}

\begin{figure}[htbp]
\centering
\includegraphics[width=\textwidth]{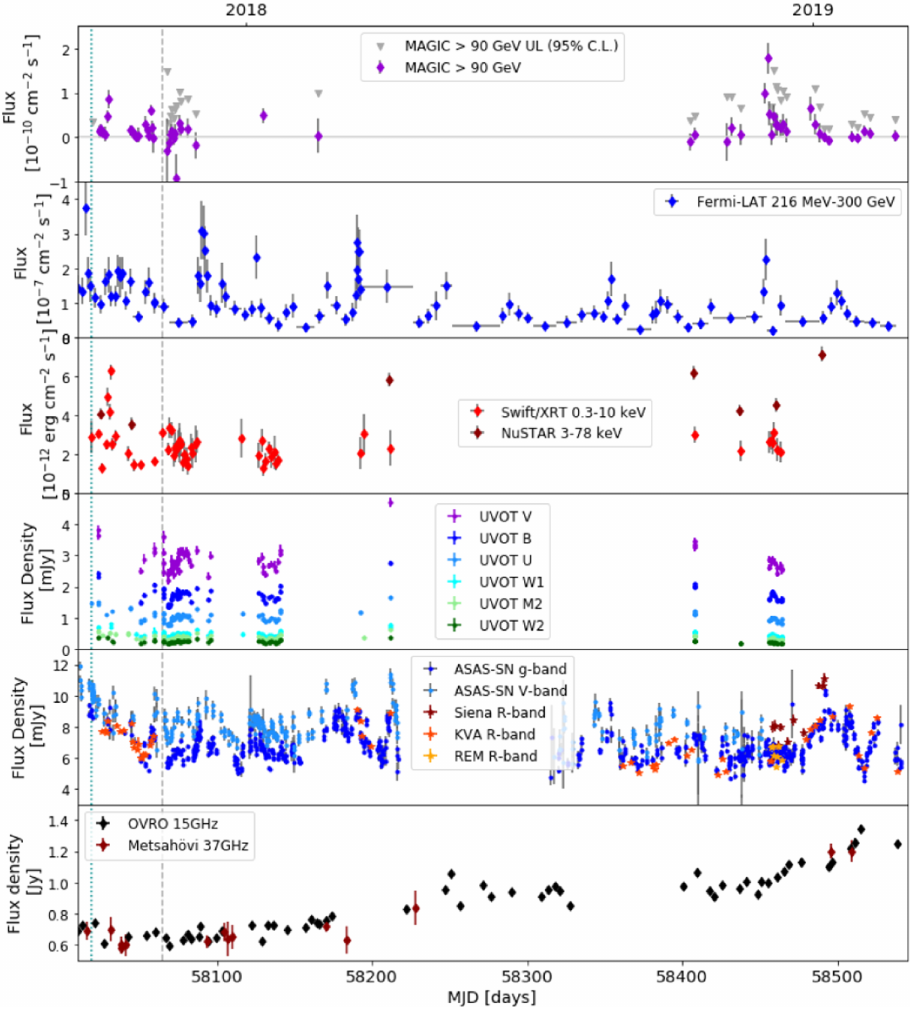}
\caption{Multi-wavelength light curve of TXS\,0506+056 collected during the 2017-2019 campaign from VHE (top panel) to radio (bottom panel). The plot includes MAGIC, \textit{Fermi}-LAT, \textit{Swift}-XRT, \textit{NuSTAR}, \textit{Swift}-UVOT, ASAS-SN, KVA, REM, Siena, OVRO and Mets{\"a}hovi data. The dotted light blue line shows the arrival time of the IceCube-170922A neutrino. Optical and X-ray fluxes are corrected for Galactic absorption, while MAGIC data are not corrected for pair-production absorption on the extragalactic background light. Image reproduced with permission from \citet{acciari22}, copyright by the author(s)}\label{0506_fig3}
\end{figure}

An extended multi-wavelength campaign has been organized between November 2017 and February 2019 to monitor TXS\,0506+056 from radio to VHE, including Mets{\"a}hovi, OVRO, ASAS-SN, KVA, REM, \textit{Swift}, \textit{NuSTAR}, \textit{Fermi}-LAT and MAGIC telescopes \citep{acciari22} (see Fig.~\ref{0506_fig3}). 

At VHE, MAGIC observations have shown a short flare on 2018 December 1--3 very similar in terms of flux level and photon index to the 2017 September flare, while just an excess at 4-$\sigma$ level has been detected during the rest of the campaign (74 hours of accumulated exposure). Unlike the long-term brightening observed in 2017, \textit{Fermi}-LAT detected a much lower average flux with several short day-to-week timescale flares. However, no significant flares are detected in optical and X-rays, with only an increasing trend of the radio flux at 15 GHz and 37 GHz with an increase by a factor of 2 throughout the campaign. This can be an indication that the radio emitting region is different from the zone emitting at higher energies. 

\begin{figure}[ht]
\centering
\includegraphics[width=0.49\textwidth]{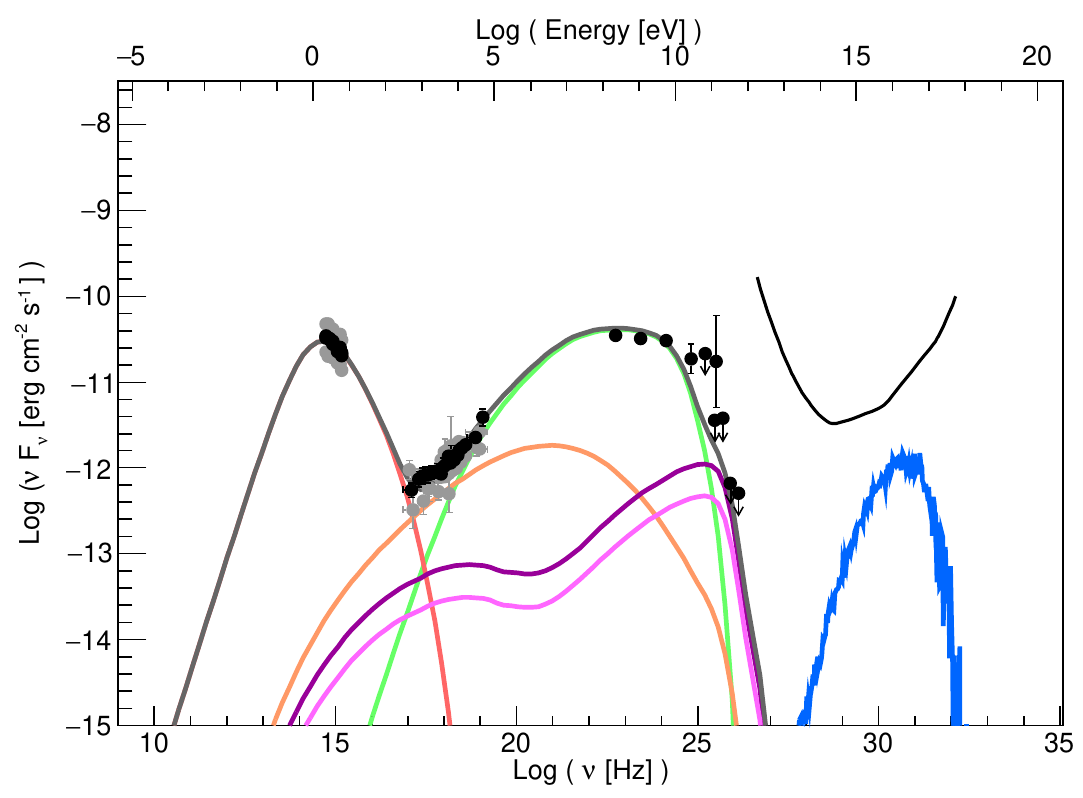}
\includegraphics[width=0.49\textwidth]{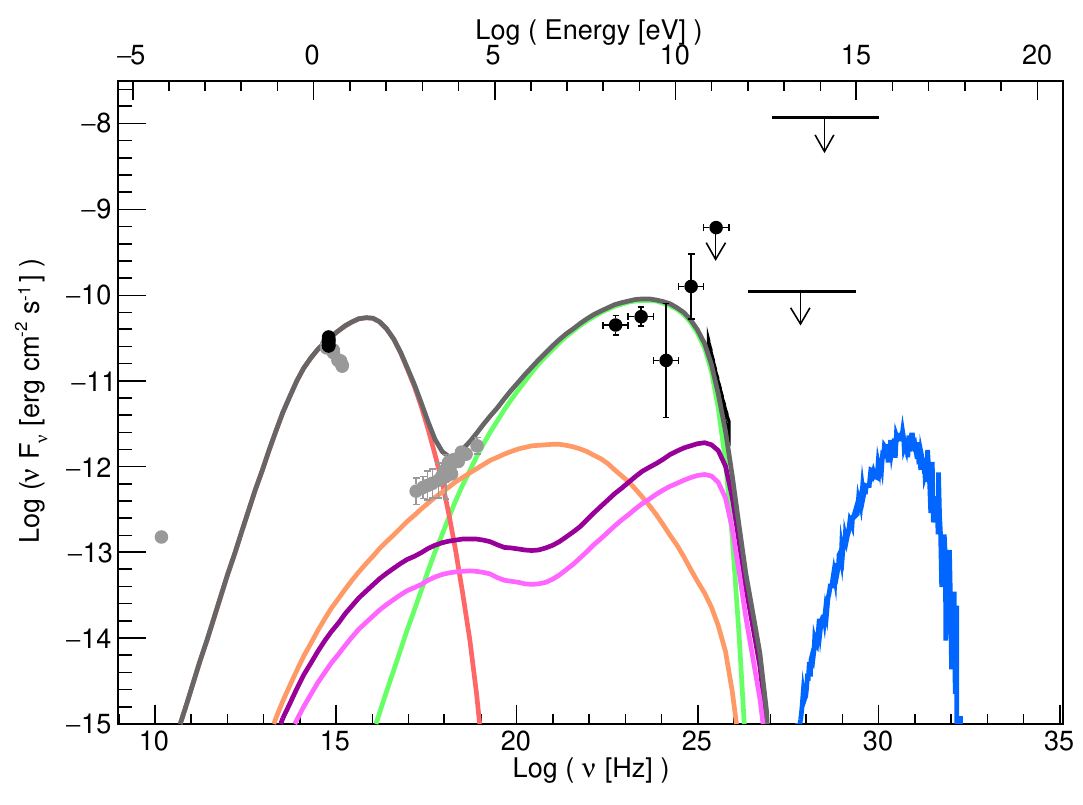}
\caption{SEDs of TXS\,0506+056 including \textit{Fermi}-LAT, MAGIC, \textit{Swift}, \textit{NuSTAR}, ASAS-SN, OVRO and Mets{\"a}hovi data collected in 2017-2019 campaign and modelled in a lepto-hadronic scenario. The SED representing the low (2018 October) and high (2018 December) state is shown on the left and right panel, respectively. The solid lines represent synchrotron emission by primary electrons (red), IC emission over photons from the jet layer (green), Bethe--Heitler cascade (orange), cascade from $\pi^{0}$ (violet) and $\pi^{\pm}$ (pink) decay, and neutrino emission (blue). In the left plot the IceCube 5-$\sigma$ sensitivity curve for point-source searches is shown, while in the right plot the IceCube and ANTARES flux upper limits (at lower and higher energies, respectively) following the detection of a VHE flare in 2018 December are shown. Image reproduced with permission from \citet{acciari22}, copyright by the author(s)}\label{0506_fig4}
\end{figure}

Only marginal spectral variability has been observed in optical-UV and $\gamma$-rays by UVOT and \textit{Fermi}-LAT, respectively, indicating that the position of the SED peaks was stable during the campaign. No spectral variability has been observed in  the hard X-ray band, as observed by \textit{NuSTAR}, while a clear harder-when-brighter behaviour was detected in the 2--10 keV energy range by \textit{Swift}-XRT. Interestingly, this behaviour disappeared by considering the 0.3--10 keV energy band, suggesting a transition between different emission mechanisms in that band. In the context of a lepto-hadronic model, this can correspond to the transition from synchrotron emission by primary electrons, through the one by Bethe--Heitler pairs, to IC emission. According to a lepto-hadronic scenario with the hadronic emission produced by Bethe--Heitler and pion-decay cascade in X-ray and VHE $\gamma$-rays, respectively, the December 2018 flare observed in the $\gamma$-ray band should be connected to a neutrino emission too brief and not bright enough to be detected by IceCube (see Fig.~\ref{0506_fig4}). 

In addition, the observation of a significant neutrino flux from NGC\,1068 presumably coming from the AGN corona (see Sect.~\ref{1068}) suggests that neutrinos can also be produced in the cores of AGN. This poses the question whether neutrino production also in TXS\,0506+056 can be associated with the core region, but the neutrino emission expected from the corona of TXS\,0506+056 is too low to describe neutrinos from this source observed by IceCube; therefore, the jet remains the preferred location for neutrino production in that source \citep{fiorillo25}.

\subsection{NGC\,1068: the first non-jetted AGN associated with high-energy neutrino emission}\label{1068}

\begin{figure}[ht]
\centering
\includegraphics[width=0.5\textwidth]{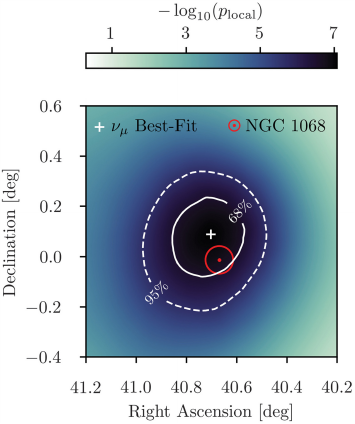}
\caption{High-resolution scan around the most significant location marked by a white cross, with contours showing its 68\% (solid) and 95\% (dashed) confidence regions. The red dot shows the position of NGC\,1068, and the red circle is its angular size in the optical wavelength. Image reproduced with permission from \citet{abbasi22b}, copyright by AAAS}\label{NGC1068_fig1}
\end{figure}

\begin{figure}[ht]
\centering
\includegraphics[width=0.9\textwidth]{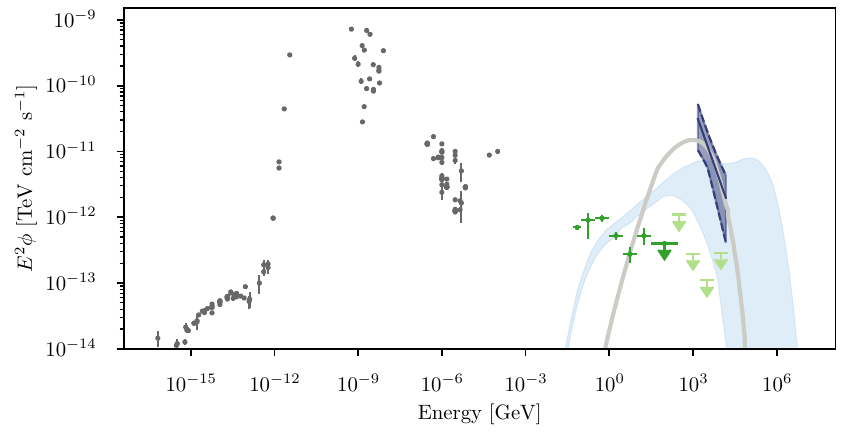}
\caption{Multi-messenger SED of NGC\,1068 where gray points show multi-frequency observations of the source from radio to X-rays. Dark and light green points indicate HE $\gamma$-ray observations from \textit{Fermi}-LAT in the 0.1--100 GeV energy range and the upper limit at VHE $\gamma$-ray observations at $E > 200\, \mathrm{GeV}$ from MAGIC \citep{acciari19}, respectively. Arrows indicate upper limits. The solid, dark blue line shows the best-fitting neutrino spectrum in the 1.5--15 TeV energy range with the dark blue shaded region indicating the 95\% confidence region. We restrict this spectrum to the range. The light blue shaded region and the gray line show the NGC\,1068 neutrino emission models from \citet{inoue20} and \citet{murase20}, respectively. The shaded region covers possible values of the gyrofactor in the between 30 and 104 used to describe uncertainty in the efficiency of the underlying particle acceleration. Adapted from \citet{abbasi22b}}\label{NGC1068_fig2}
\end{figure}

In 2022, after having analyzed 10 years of data for a sample of sources, the IceCube Collaboration reported 4.2-$\sigma$ evidence for an excess of TeV neutrinos from NGC\,1068 \citep[see Fig.~\ref{NGC1068_fig1};][]{abbasi22b}. NGC\,1068 is a nearby Seyfert II galaxy with a Compton thick AGN, significant starburst activity, and outflows. NGC\,1068 is also the most luminous Seyfert II galaxy detected in $\gamma$ rays by \textit{Fermi}-LAT so far \citep{ajello22}. The first tentative detection of neutrinos from the source was reported in \citet{aartsen20}, with a significance of 2.9-$\sigma$. The confirmation of the connection between neutrino emission and NGC\,1068 together with the observed flux of neutrino events in the 1--50 TeV energy range more than one order of magnitude higher than the $\gamma$-ray emission level observed by \textit{Fermi}-LAT and MAGIC (Fig.~\ref{NGC1068_fig2}) forced us to review the theoretical models that explain the SED of non-jetted AGN \citep[e.g.][]{inoue20,murase20b}. The suppression of a strong $\gamma$-ray emission can be related to the fact that neutrino can escape from the emitting region without interactions, while photons can be absorbed during additional interaction on other photons and depending on the environment the connection observed between neutrino and $\gamma$-rays can be significantly different. 

In principle, several regions can correspond to the potential site of cosmic-ray acceleration and neutrino production processes in this source: the starburst region in the spiral arms of the host galaxy, the kpc jet, a sub-kpc molecular outflow and the accretion disc-corona. Unfortunately, the current resolution of the IceCube detector does not allow the identification of the neutrino emitting region, thus the best way to identify the region in which the neutrino are produced in NGC\,1068 is by means of theoretical modelling of its SED. Three different theoretical scenarios have been proposed: diffusive shock acceleration at shock waves, stochastic acceleration in magneto-hydrodynamic turbulence, and systematic acceleration in magnetic reconnection.
\citet{fang23} proposed the inner dusty torus and the jet-ISM interacting region as a potential neutrino production site. Neutrino production in the torus has already been investigated in \citet{inoue22} resulting in a low neutrino flux. The model proposed by \citet{fang23} is focused on the inner part of the torus, where the jet interacts with the ISM, and not with the outer torus region, with a radiation field extended up to UV and a higher magnetic field. However, measurement of the intrinsic power of jets in models favoring jet-ISM interactions is frequently criticized. This is primarily due to the uncertainty surrounding energy transfer efficiency and the reliance on radio luminosity as a proxy for kinetic power (Fig.~\ref{NGC1068_fig2}).

\begin{figure}[ht]
\centering
\includegraphics[width=\textwidth]{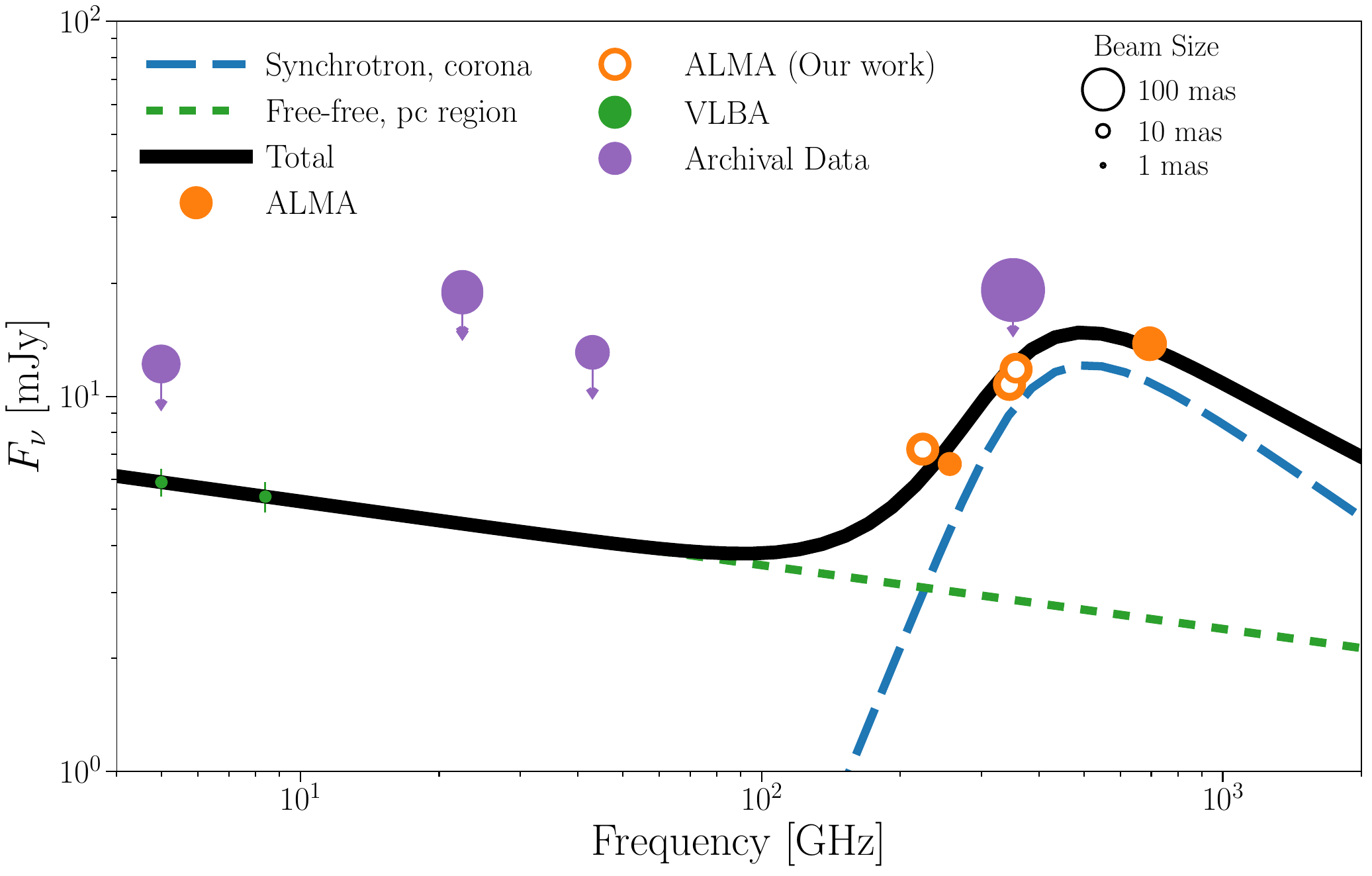}
\caption{Spectrum of NGC\,1068 in the cm-mm band. The data points in green are from VLBA \citep{gallimore04}, while those in orange from ALMA \citep{garciaburillo16,garciaburillo19,impellizzeri19}. The open points are ALMA data analyzed in \citet{inoue20}. The size of the circles is the beam size. The blue-dashed and green-dotted lines are the coronal synchrotron and free-free emission components, respectively. The black solid line is the sum of the two components. Image reproduced with permission from \citet{inoue20}, copyright by AAS}\label{NGC1068_fig3}
\end{figure}

Among the different possible regions, the accretion disc-corona is the favourable one, in agreement with the results of model-independent studies on the connection between neutrino and $\gamma$-rays in this source for which the neutrino is most likely produced in a region within $\sim$30--100 Schwarzschild radii, in particular if produced by photo-meson processes \citep{murase22}. In fact, the disc-corona system has an adequate density of X-ray photons to provide the targets for protons needed for neutrino production and sufficient density of optical and infrared photons to absorb the associated emission that should be emitted but is not observed in $\gamma$-rays \citep[e.g.,][]{padovani24}. In this context, the corona super-hot plasma around the black hole can accelerate protons, carrying a few percent of the thermal energy through plasma turbulence or shock acceleration, leading to the creation of neutrinos and $\gamma$-rays (see e.g. the models reported in Fig.~\ref{NGC1068_fig2}). Investigating the corona activity in the millimeter band with ALMA, \citet{inoue20} found that non-thermal emission from the corona with coronal parameter (i.e magnetic field strength and corona size) similar to those inferred from other nearby Seyfert can be consistent with a spectral excess observed in the mm band, providing independent electromagnetic support for the coronal model as the main mechanism for producing neutrinos in NGC\,1068. In fact, the extrapolation of the emission at cm, in which a homogeneous emission due to free-free emission has been detected with VLBA observations by \citet{gallimore04}, is not consistent with the flux observed at higher frequencies, with the spectral shape of the excess consistent with non-thermal emission from a compact region with a size compatible with the corona (Fig.~\ref{NGC1068_fig3}). Based on 3D particle-in-cell simulations of relativistic reconnection, \citet{fiorillo24} proposed that the acceleration mechanism of the non-thermal protons may be due to a large-scale magnetic reconnection layer with a size of a few Schwarzschild radii. In the proposed scenario, an energy equipartition between magnetic fields, hard X-ray photons from the corona, and non-thermal protons in the reconnection region is expected. This scenario can also explain part of the diffuse neutrino emission. In case of a non-negligible ($\sim$10\%) fraction of AGN coronae with a high plasma magnetization (i.e. $\sigma_{p}$ $\sim$ 10$^{5}$), the coronal emission from AGN is proposed to explain the diffuse neutrino flux observed by IceCube up to $\sim$1 PeV, while fails at higher energies due to the strong Bethe-Heitler proton cooling and an additional contribution, e.g. from a jetted AGN population, is suggested \citep{karavola26}.

\begin{figure}[ht]
\centering
\includegraphics[width=0.49\textwidth]{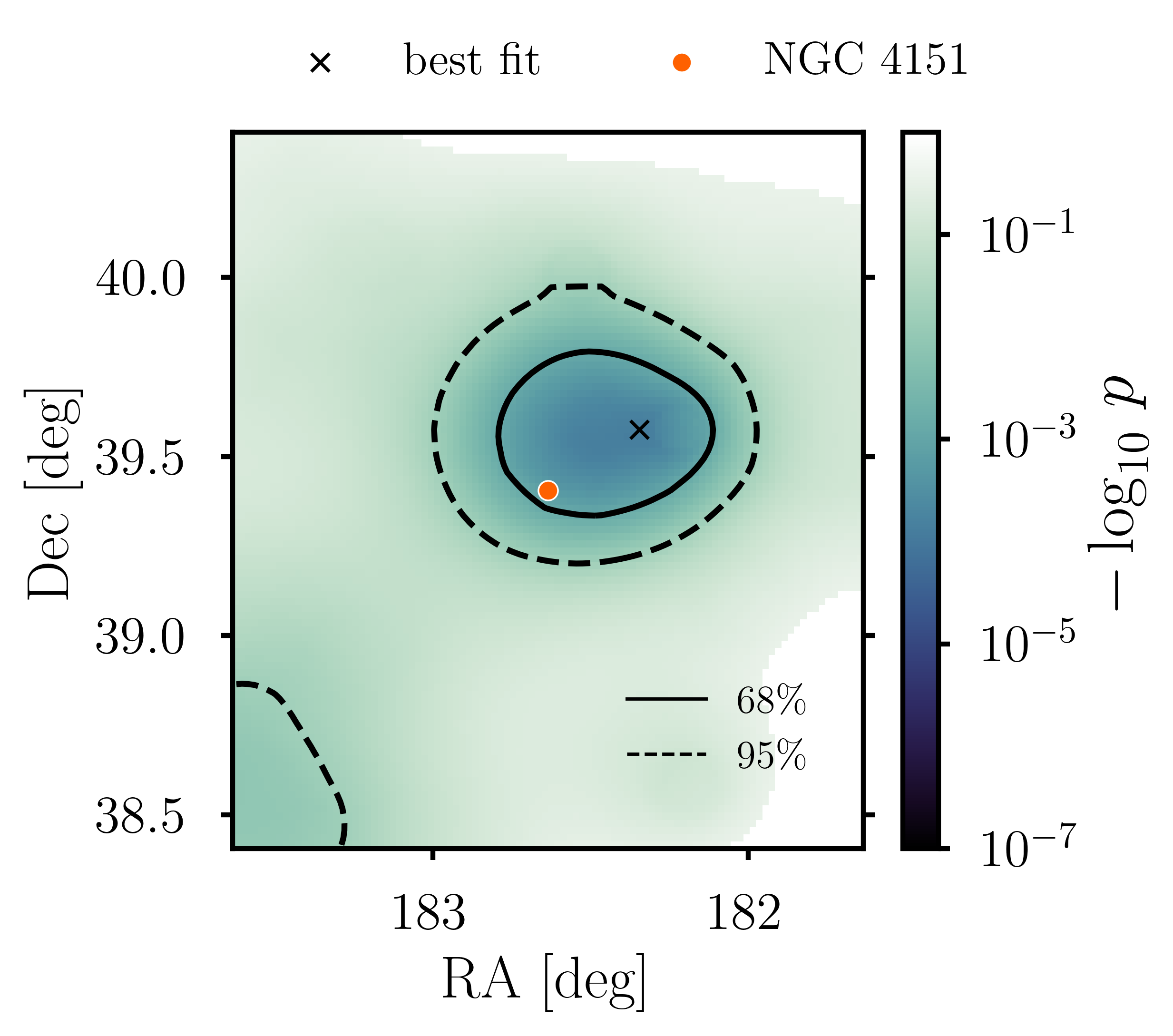}
\includegraphics[width=0.49\textwidth]{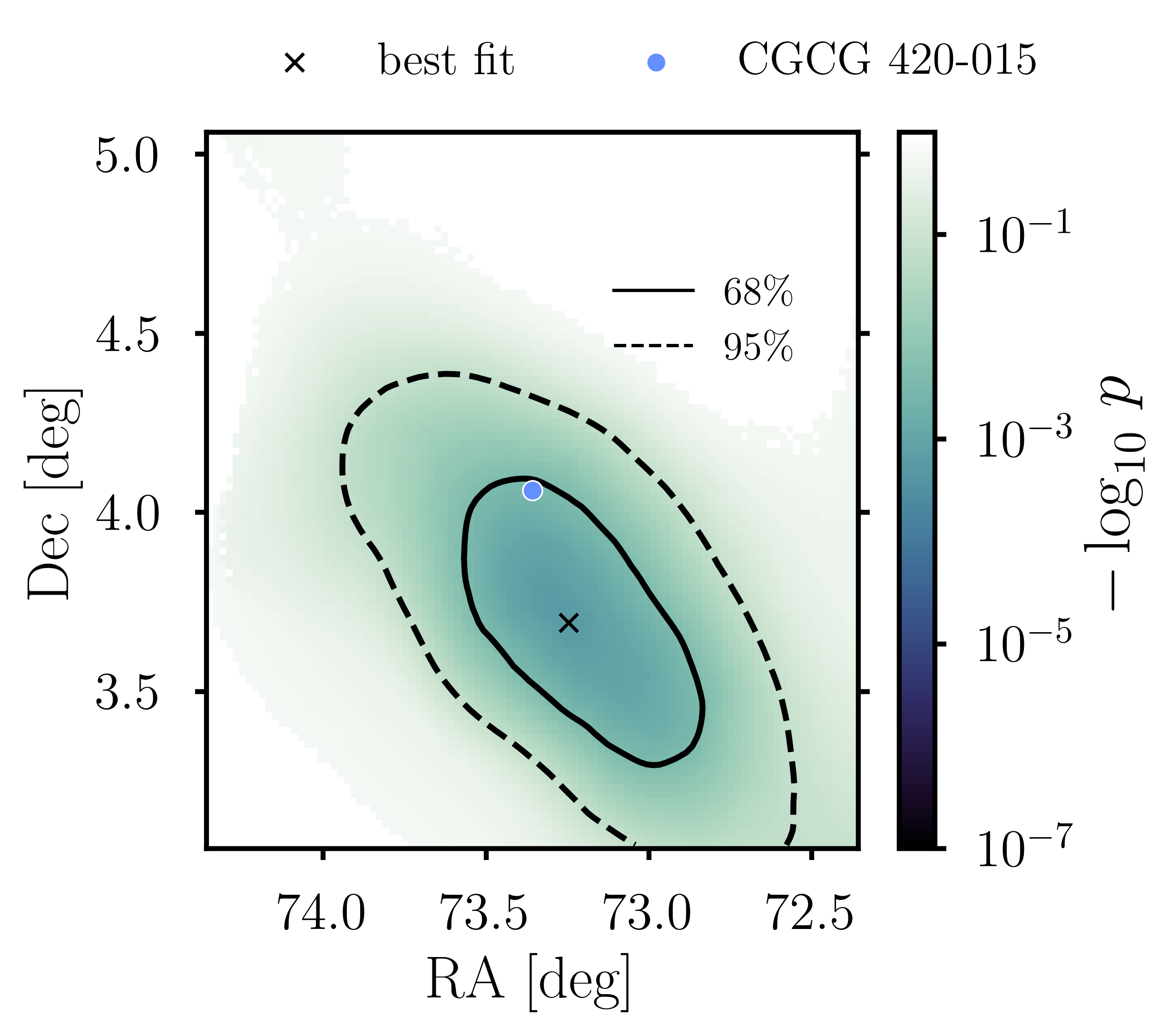}
\caption{Local (pre-trial) p-value maps around NGC\,4151 {\it (left panel)}, and CGCG 420-015 {\it (right panel)} with the disc-corona model fit. Colored points show the locations of sources and crosses show the best-fit locations. Contours correspond to 68\% (solid) and 95\% (dashed) confidence. Image reproduced with permission from \citet{abbasi25}, copyright by the author(s)}\label{NGC1068_fig4}
\end{figure}

\begin{figure}[ht]
\centering
\includegraphics[width=\textwidth]{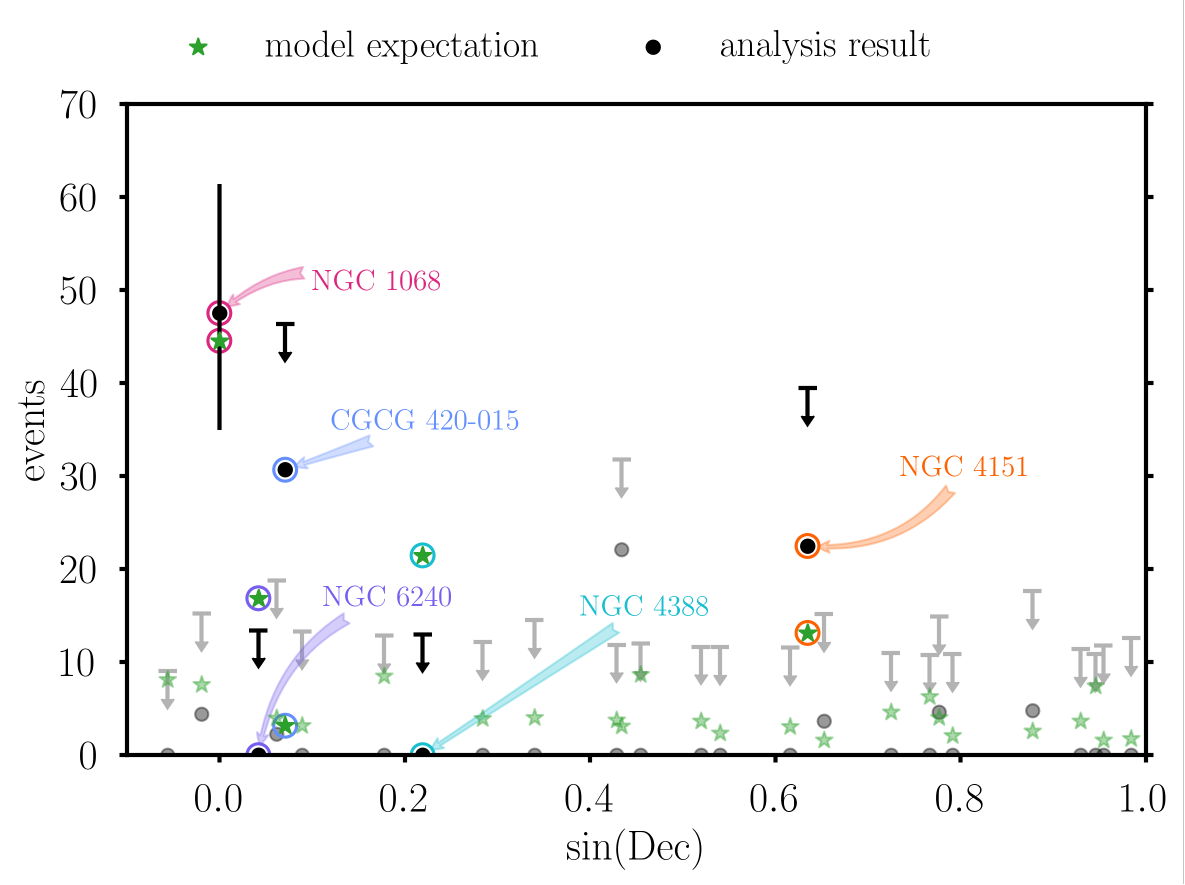}
\caption{Based on the disc-corona model presented in \citet{abbasi25}, the expected numbers of events (green stars) from the model and the best-fitted numbers of signal events (black circles) for individual sources are shown, with the five candidate sources with the strongest expected neutrino signal highlighted. The 90\% confidence level upper limits are shown as down arrows. Image reproduced with permission from \citet{abbasi25}, copyright by the author(s)}\label{NGC1068_fig5}
\end{figure}

In 2025, \citet{abbasi25} reported significant new evidence identifying X-ray bright Seyfert galaxies as a potential key class of high-energy neutrino emitting sources. Being the central core of these AGN obscured to $\gamma$ rays but bright in X-rays makes them prime candidates for neutrino production. Based on disc-corona models, which predict that neutrinos are produced in the hot, magnetized corona surrounding the central SMBH, neutrino emission can be correlated with X-rays because they are tracers of coronal activity. A binomial test of 27 northern-sky Seyfert galaxies (excluding NGC\,1068) found a combined significance of 2.7-$\sigma$, suggesting that a subset of these galaxies share similar physical properties with NGC\,1068. Beyond NGC\,1068, the analyses identified neutrino excesses associated with NGC\,4151 and CGCG 420-015 (Fig.~\ref{NGC1068_fig4}). Other two sources, NGC\,6240 and NGC\,4388, appear to be ruled out by the model prediction proposed in \citet{abbasi25}, based on stochastic particle acceleration and high cosmic-ray pressure in the AGN corona, because the 90\% confidence level upper limits are below the expected fluxes for these two candidate sources (Fig.~\ref{NGC1068_fig5}). Moreover, current data suggest that if all Seyfert galaxies were as efficient as NGC\,1068, the observed diffuse neutrino flux would be much higher, indicating that NGC\,1068 may be exceptionally powerful. 

A potential correlation between the unabsorbed 15--55 keV and neutrino luminosities (fluxes) of four Seyfert (NGC\,1068, NGC\,4151, CGCG\,420$-$015, and NGC\,3079) and two blazars (TXS 0506$+$056 and GB6\,J1542$+$6129) has been suggested, implying that the same mechanism produces neutrinos in Seyfert and blazars, not involving a strong jet contribution \citep{kun24}. However, the sample size is small and, at least in case of TXS\,0506$+$056, \citet{fiorillo25} have shown that, although theoretically possible under certain extreme conditions, coronal neutrino emission from this source is too low to describe the neutrino emission observed by IceCube, with the jet that remains the preferred location for neutrino production. 

Furthermore, a search of neutrino emission from 14 bright Seyfert galaxies in the Southern Hemisphere based on a set of muon neutrinos with interaction vertices inside the IceCube detector does not identify significant neutrino emission from individual sources. However, assuming the disc-corona model spectrum, the stacking analysis of all sources results in a signal inconsistent with the background at 3-$\sigma$ level of significance, driven by three sources: Circinus Galaxy, NGC\,7582, and ESO\,138-1 \citep{abbasi26}.

Finally, the cosmic-rays possibly associated with TXS\,0506+056 are of higher energies ($\sim$ 800 TeV\,--\,90 PeV) with respect to the energies associated to NGC\,1068 ($\sim$ 40--400 TeV), suggesting that jetted and non-jetted AGN can contribute to different parts of the neutrino astrophysical background emission.


\section{Potential associations between AGN and high-energy neutrinos}\label{sec3}

In addition to TXS\,0506+056 and NGC\,1068, other possible associations between high-energy neutrinos and AGN have been reported, although less significant \citep{garrappa19, franckowiak20}. Among them, the candidates most thoroughly investigated in the literature are: PKS\,1502+106, PKS\,1424$-$418, BZB\,J0955+3551, and PKS\,0735+178. In the rest of the section we discuss observational evidence and related theoretical models proposed for these four sources. Other candidates of potential interest, not discussed in detail here, are: PKS\,0239$+$108 \citep[also known as J0242+1101;][]{albert24}, and PKS\,0446$+$11 \citep{kovalev26}.

\subsection{PKS\,1502+106}

\begin{figure}[ht]
\centering
\includegraphics[width=\textwidth]{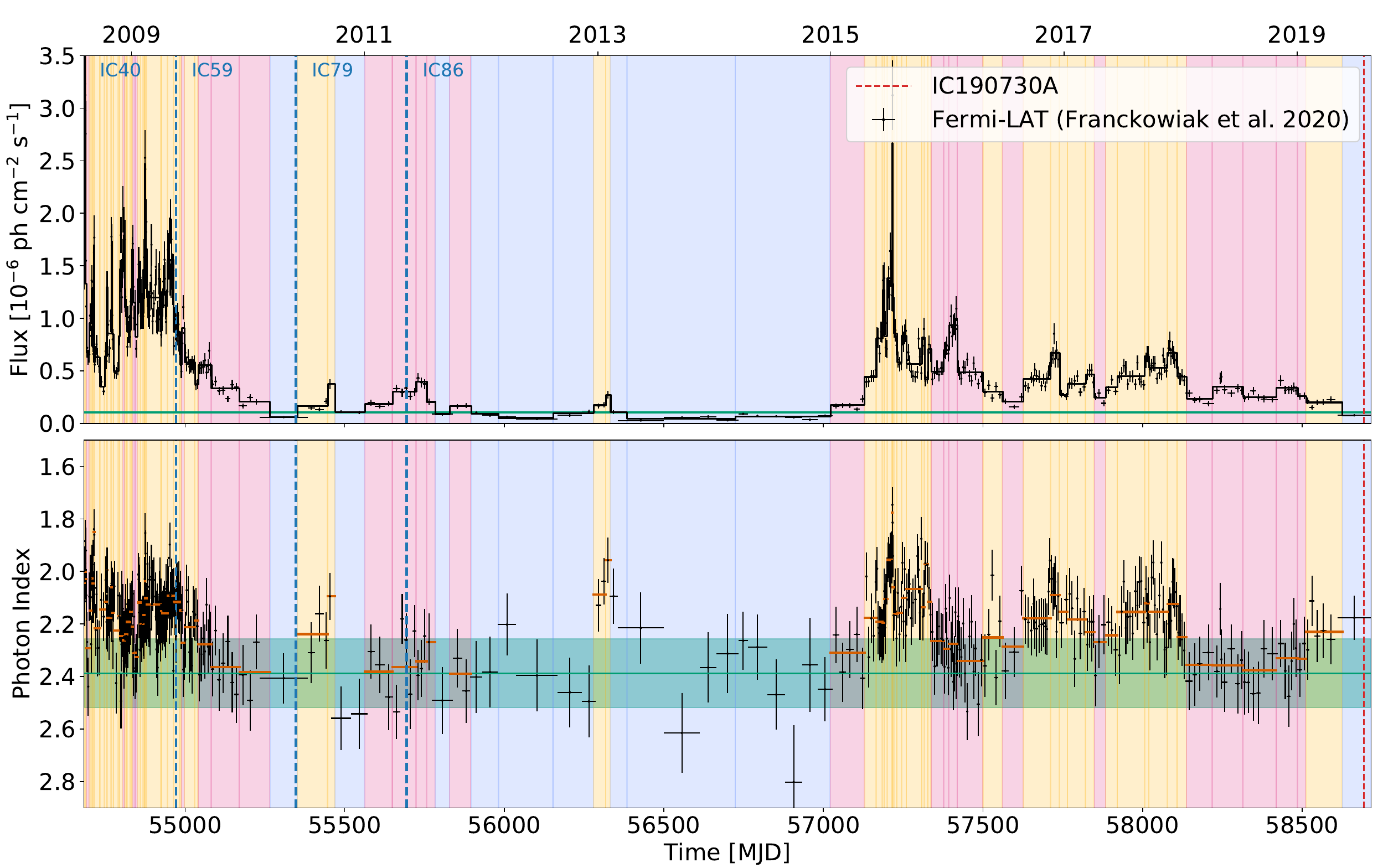}
\caption{{\it Top:} \textit{Fermi}-LAT light curve of PKS\,1502+106 during 11 years of observations taken from \citet{franckowiak20}, divided in the three periods selected for the SED modeling: quiescent (blue), flaring with hard spectrum (yellow), flaring with soft spectrum (pink) spectrum. {\it Bottom:} Photon index collected by \textit{Fermi}-LAT in the same period of the light curve. Image reproduced with permission from \citet{rodrigues21}, copyright by AAS}\label{1502_fig1}
\end{figure}

In 2019 July, IceCube detected a 300 TeV high-energy neutrino from the direction of the powerful FSRQ PKS\,1502+106, with a high astrophysical neutrino signalness.\footnote{\url{https://gcn.nasa.gov/circulars/25225}} This is one of the brightest and highly variable $\gamma$-ray sources in the fourth catalog of AGN detected by \textit{Fermi}-LAT \citep{ajello22}. Due to the large redshift \citep[$z$ = 1.84;][]{hewett10}, this high flux results in a very high luminosity. The source is highly variable at radio frequencies with a one-sided, core-dominated, curved jet, as observed in VLBI images \citep{an04}. Unlike the 2017 neutrino event related to TXS\,0506+056, no significant $\gamma$-ray flaring activity has been observed from PKS\,1502+106 at the time of the neutrino event IC-190730A \citep{franckowiak20} (see Fig.~\ref{1502_fig1}). In contrast, a long-term radio outburst has been observed from this source that started in 2014 and reached the peak of activity at the time of the IC-190730A event \citep{kiehlmann19}. 

\begin{figure}[ht]
\centering
\includegraphics[width=\textwidth]{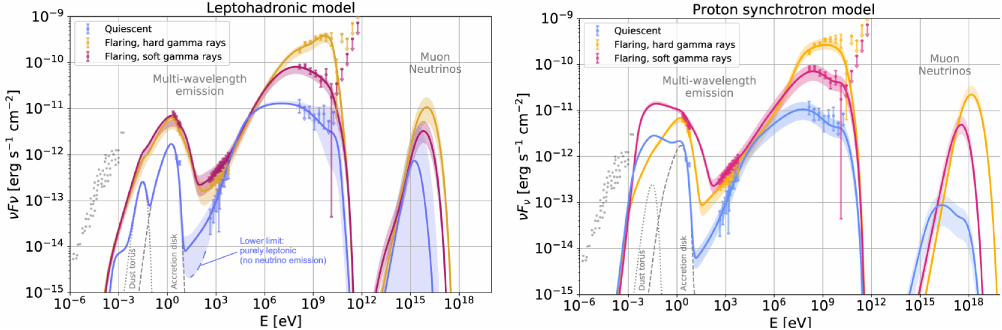}
\caption{The predicted SED and neutrino spectra from PKS\,1502+106 obtained with the lepto-hadronic model (left panel) and the proton synchrotron model (right panel) with three different parameter sets. The shaded areas correspond to the uncertainty in the non-thermal proton power. The multi-wavelength data selected in the three periods reported in Fig. \ref{1502_fig1} are reported as color data points, while archival radio data are reported as grey points. Adapted from \citet{rodrigues21}}\label{1502_fig2}
\end{figure}

A detailed multi-wavelength and multi-messenger modeling of the source has been reported in \citet{rodrigues21}. Three different activity states of the blazar, one quiescent state and two flaring states with hard and soft $\gamma$-ray spectra, have been selected to be tested with two different theoretical models. In these two models, the X-ray to $\gamma$-ray part of the SED is represented by a lepto-hadronic model with photohadronic processes and a proton synchrotron component, respectively (Fig.~\ref{1502_fig2}). The emission can be well described by a hadronic contribution, both during quiescent and flaring states. A purely leptonic one-zone model does not adequately explain the X-ray fluxes during flaring states, with a subdominant contribution due to electromagnetic cascades initiated by hadronic interactions, which can be interpreted as additional evidence for cosmic-ray acceleration in the source independently of neutrino observations. Different from the compact region responsible for the optical-to-$\gamma$-ray emission, a larger dissipation region has been proposed for the radio emission. In this context, the bright radio activity state observed at the time of the neutrino emission seems to suggest a common origin, but such a scenario is not feasible in a one-zone model within the theoretical scenario proposed in that work, suggesting more complex geometries to be investigated. 

\citet{oikonomou21} proposed a lepto-hadronic model applied not to multi-wavelength data collected contemporaneously to the neutrino detection but during an average long-term state of the source in which the SED is built with \textit{WISE}, \textit{Swift} (UVOT and XRT), and \textit{Fermi}-LAT data. An emitting region beyond the BLR and inside the dust torus produces the scenario most consistent with the observations. It is interesting to note that in this model the required proton luminosity is consistent with the average required proton luminosity if blazars power the observed ultra-high-energy cosmic rays (UHECR) ﬂux and well below the Eddington luminosity of the source. 

\subsection{PKS\,1424$-$418}

\begin{figure}[ht]
\centering
\includegraphics[width=\textwidth]{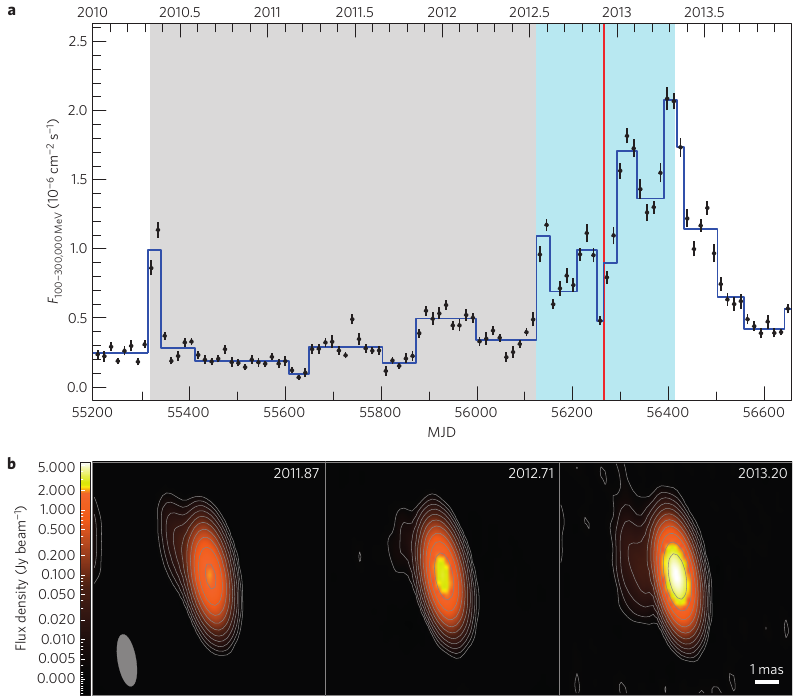}
\caption{{\it Top panel}: \textit{Fermi}-LAT light curve of PKS\,1424$-$418 collected during 2010--2014 with 2-week time-bins in the 0.1--300 GeV energy range is shown in black, while the Bayesian block light curve in blue. The red line marks the detection of the IC35 neutrino. {\it Botton panel}: VLBI images of the source obtained with TANAMI at 8.4 GHz on 2011 November, 2012 September and 2013 March in uniform color scale. All contours start at 3.3 mJy/beam, increasing logarithmically by factors of 2. The images were convolved with the beam from all three observations shown in the bottom left. Image reproduced with permission from \citet{kadler16}, copyright by Macmillan}\label{1424_fig1}
\end{figure}

\textit{Fermi}-LAT has observed long-lasting strong $\gamma$-ray flaring activity from the FSRQ PKS\,1424$-$418 claimed in temporal and positional concordance with the third event neutrino (IC35) detected at 2 PeV by IceCube on 2012 December 4 \citep{kadler16}. In addition to the $\gamma$-ray increase in activity, the radio flux density of the radio core increases from $\sim$1.5 Jy on 2011 November 13 to $\sim$6 Jy on 2013 March 14 (see Fig.~\ref{1424_fig1}), possibly marking the ejection of a new jet component.

However, the association between the neutrino and PKS\,1424$-$418 presents several open questions and problems. Uncertainty about the exact direction of arrival of the neutrino makes it difficult to unambiguously identify PKS\,1424$-$418 as its source, allowing for the possibility of random coincidences with other celestial sources in the same area. Due to the large positional uncertainty (a median positional uncertainty of $R_{50}$ = 15.9$^{\circ}$), several $\gamma$-ray-emitting AGN are found in the $R_{50}$ field of IC35 that includes 17 blazars, two radio galaxies, and one starburst galaxy. The closest AGN and brightest radio source in the field was Cen A, which has been excluded as a potential neutrino emitter because the radio emission of Cen A is emitted from the kpc scale lobes of this FR I radio galaxy and the extrapolated SED at PeV energies is too low \citep{padovani14}. Although the probability of random coincidence has been estimated at around 5\% in \citet{kadler16}, a single event is not sufficient for definitive and robust confirmation. Identifying neutrino sources usually requires an analysis that integrates longer periods or a larger number of correlated events. In addition, follow-up analyses of the $\gamma$-ray light curve revealed that the actual neutrino detection coincided with a local minimum of the $\gamma$-ray flux rather than with the peak of the flare. This raises questions about physical models that assume a direct and simultaneous correlation between $\gamma$-ray emission and neutrino production. 

\begin{figure}[ht]
\centering
\includegraphics[width=0.8\textwidth]{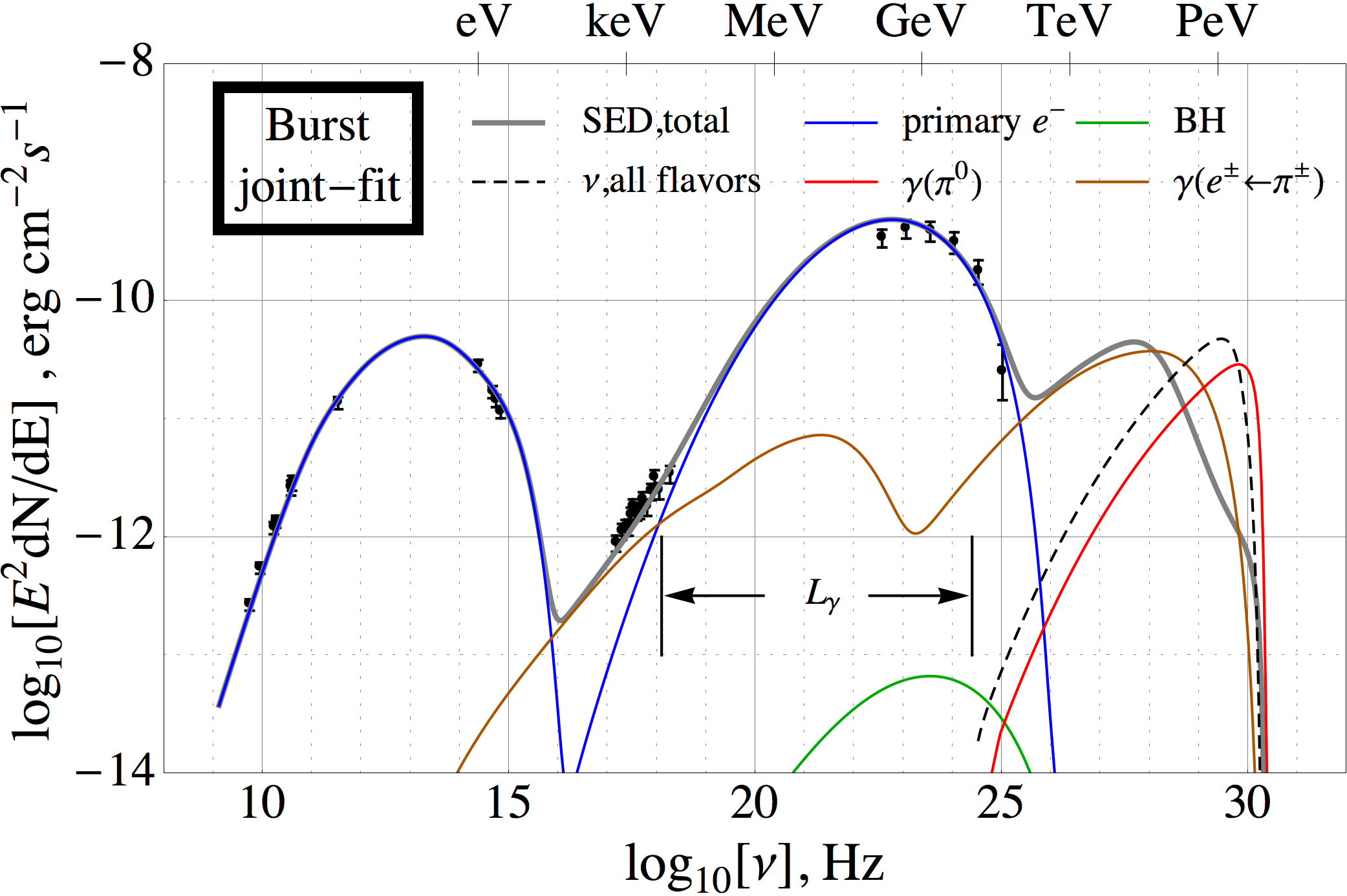}
\caption{SED of PKS\,1424$-418$ collected by \citet{kadler16} at the time of the neutrino detection. Gray curve: total SED; blue curve: emission from primary; green curve: emission from pairs generated via the black hole process; brown curve: $\gamma$-ray injection spectrum from pairs via decay; red curve: $\gamma$-ray injection spectrum from decay; black dashed curve: neutrino, all flavors. Image reproduced with permission from \citet{gao17}, copyright by AAS}\label{1424_fig2}
\end{figure}

Some alternative theoretical scenarios suggest that the neutrino could be produced in regions temporarily obscured by $\gamma$-rays, but this complicates the simple interpretation of a common origin \citep{kun21}, with leptonic (electron-driven) and hadronic (proton-driven) processes decouple depending on the target photon field's opacity. Moreover, a calorimetric estimate of the expected neutrino flux found a very small probability of the detection of a neutrino by IceCube during a multi-wavelength flare of PKS\,1424$-$418 \citep{kreter20}. Finally, purely hadronic models, which predict a strong direct correlation between $\gamma$-ray emission and neutrinos, fail to adequately describe the complete SED of PKS\,1424$-$418. Lepto-hadronic models, which include both leptonic (electron/positron) and hadronic (proton) components, are more effective, but indicate that the hadronic (neutrino-producing) component may be subdominant, weakening the certainty of direct association \citep{gao17} (see Fig.~\ref{1424_fig2}). 

\subsection{BZB\,J0955+3551}

On 2020 January 7, the IceCube Collaboration reported the detection of a high-energy neutrino detection of possible astrophysical origin. IceCube announced the detection of two additional neutrino candidates in spatial coincidence with the 90\% containment region of IC-200107A in a time range of two days around the alert. The blazar BZB J0955+3551 (also known as 3HSP\, J095507.9+355101) has been suggested to be spatially coincident with the neutrino event IC-200107A detected by IceCube. In particular, although the source has not been significantly detected by \textit{Fermi}-LAT at the time of neutrino arrival or prior on a monthly/year time-scale \citep{garrappa20}, \textit{Swift}-XRT has measured a strong X-ray flare the day after the neutrino detection \citep{krauss20}. Another $\gamma$-ray source included in the 4FGL catalog lies within the 90\% positional uncertainty of IC-200107A, but no contemporaneous flux enhancement in X-ray or $\gamma$-ray band has been observed \citep{krauss20, garrappa20}, making BZB\,J0955+3551 the best candidate. 

\begin{figure}[ht]
\centering
\includegraphics[width=\textwidth]{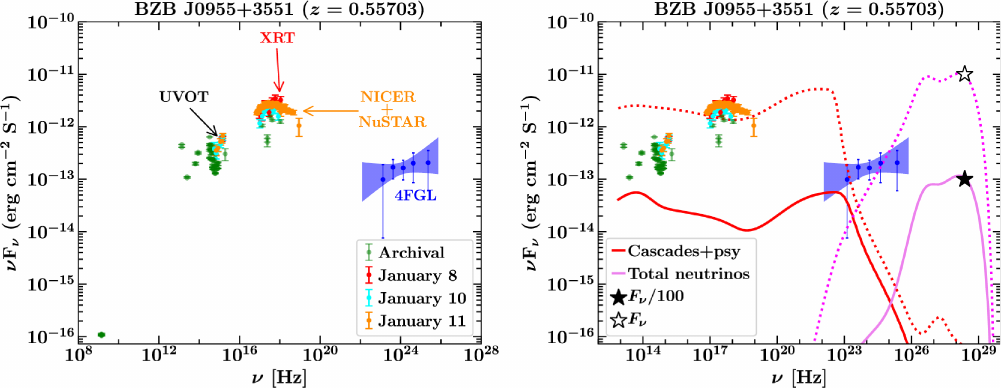}
\caption{{\it Left}: The SED of BZB\,J0955+3551 obtained with \textit{Swift}, \textit{NICER}, and \textit{NuSTAR} data collected during 2020 January 8, 10, and 11 together with the average $\gamma$-ray spectrum of the source reported in the 4FGL catalog. {\it Right}: at the SED shown on the left, the results of hadronic simulation obtained from the optical-to-X-ray constraint is superimposed. The expected 100 TeV neutrino flux (black open star) has been divided by a factor of 100 (black filled star) to take into account the Eddington bias. Adapted from \citet{paliya20}}\label{0955_fig1}
\end{figure}

BZB J0955+3551 belongs to the class of extreme blazars and, at the time of the neutrino detection, \textit{Swift}-XRT has observed the source in a very hard state with a synchrotron peak at frequencies higher than 2 $\times$ 10$^{18}$ Hz \citep{giommi20}. Differently from the case of TXS\,0506+056, this source does not seem to be a ``masquerade BL Lac''\footnote{An intrinsically flat spectrum radio quasar with emission lines significantly diluted by a strong Doppler-boosted jet emission.}. Based on energetic requirements constrained by the UV and X-ray flux observed on 2020 January 8, \citet{paliya20} explored two possible origins of the photon field that can produce the photo-pion emission: outside the jet or electron synchrotron, with the first one favoured. However, the relativistic proton population tentatively responsible for the neutrino emission can provide only a sub-dominant contribution to the IR-to-$\gamma$-ray emission observed, making this scenario unlikely (see Fig.~\ref{0955_fig1}). In addition, rapid variability has been observed in X-rays by \textit{NICER} and \textit{NuSTAR}, suggesting that the region producing the neutrino should not be the same as that producing the X-ray emission and that the neutrino and X-ray flare are not physically connected. 

\begin{figure}[ht]
\centering
\includegraphics[width=0.49\textwidth]{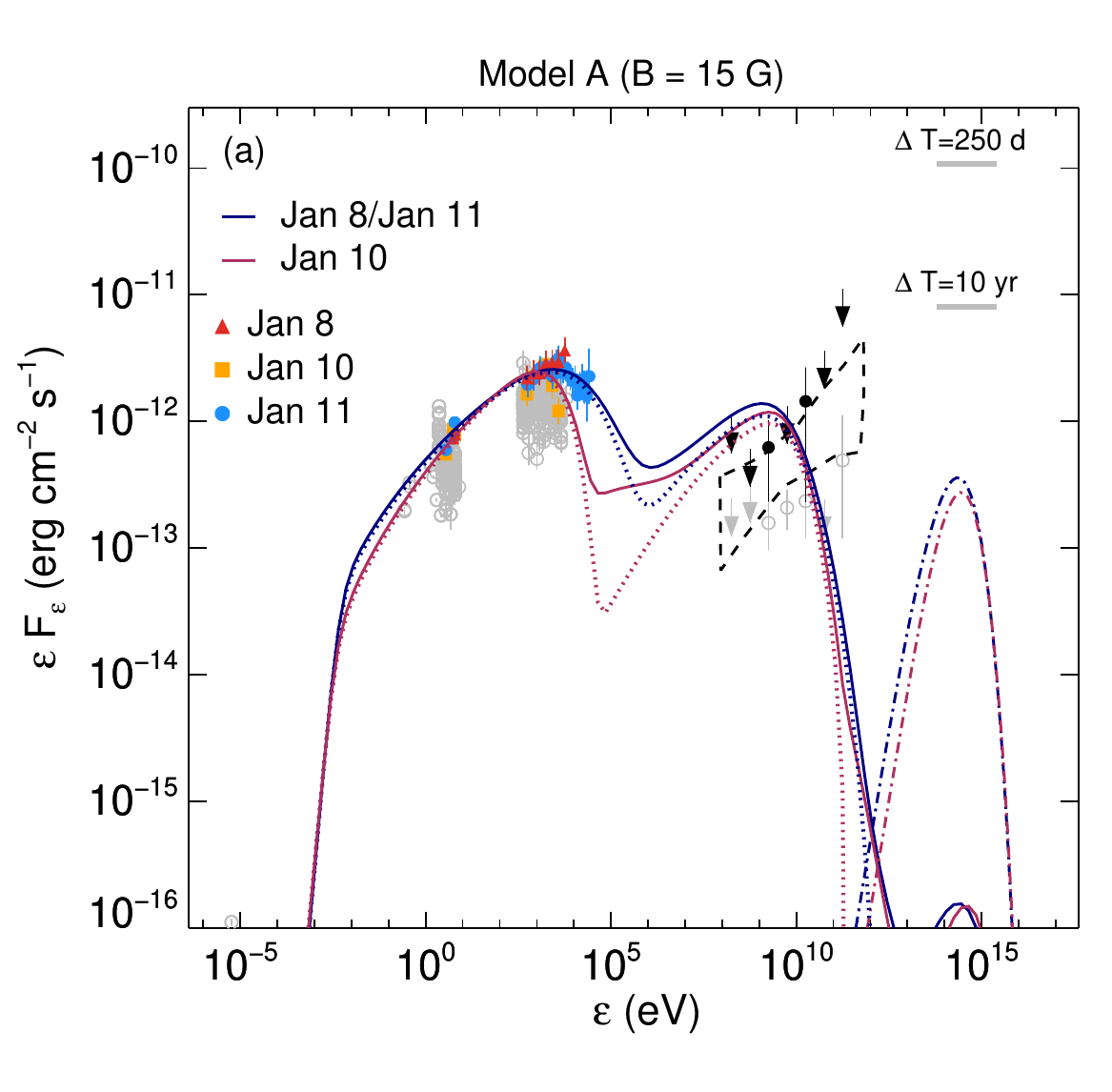}
\includegraphics[width=0.49\textwidth]{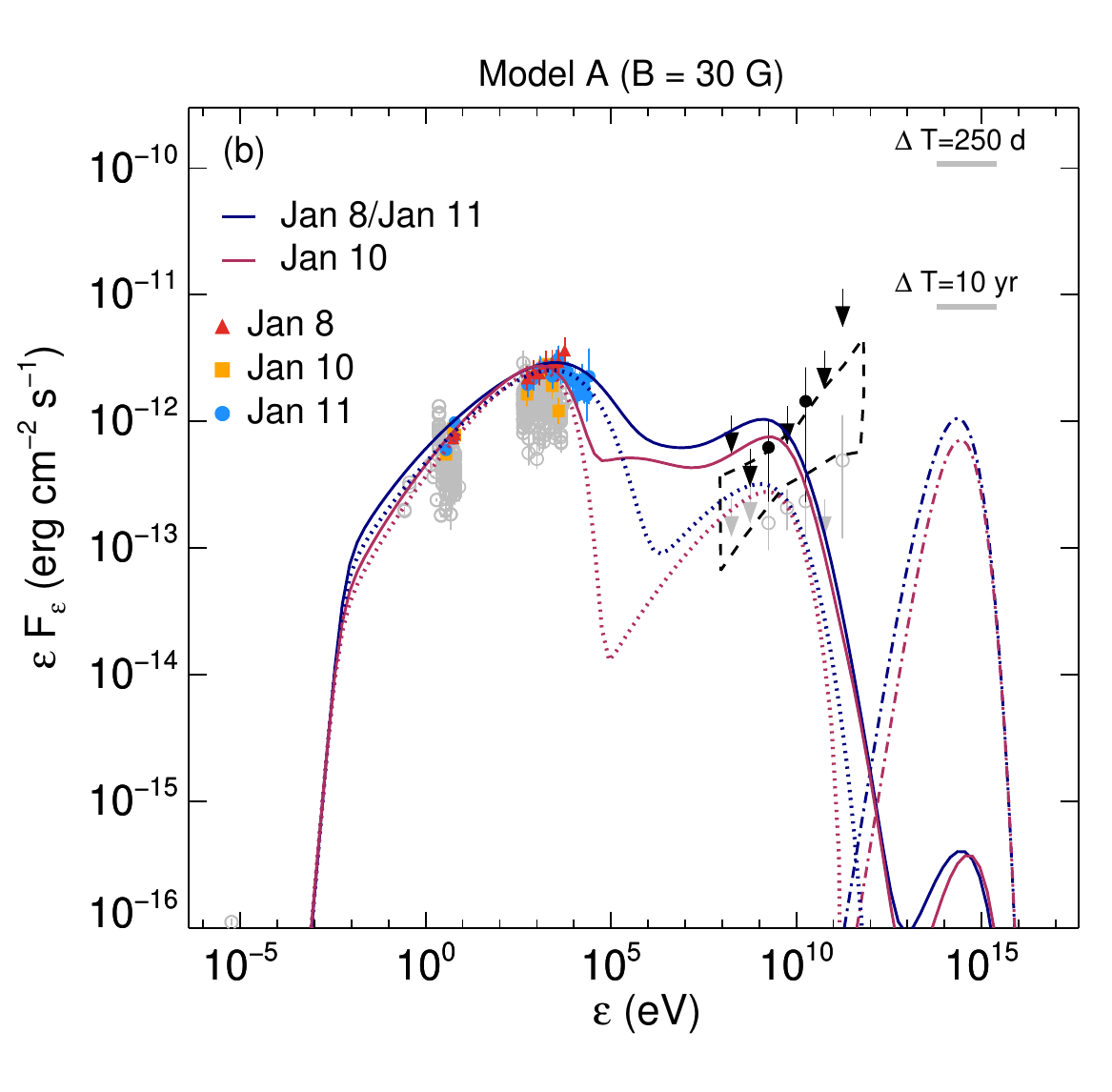}
\includegraphics[width=0.49\textwidth]{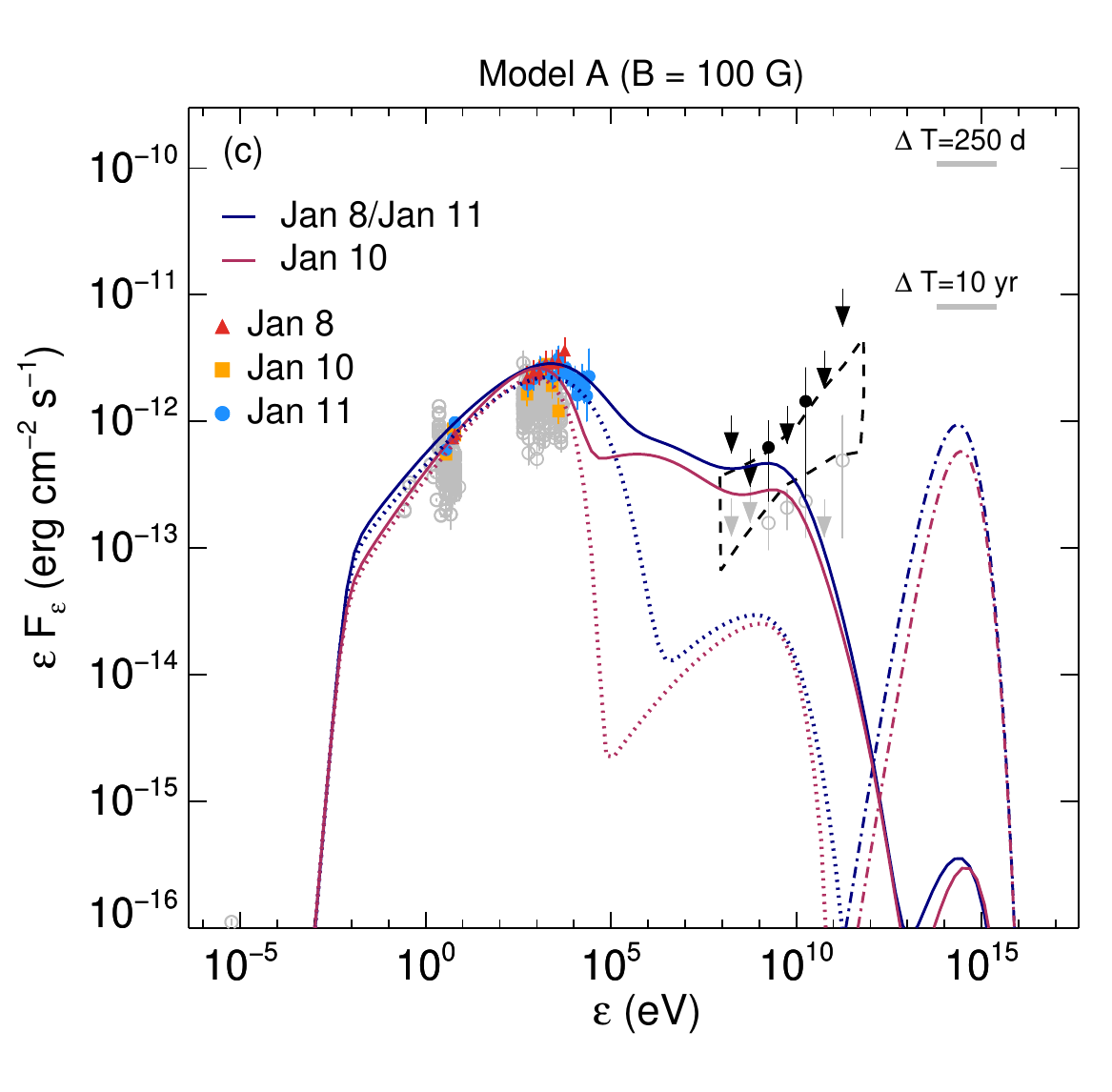}
\includegraphics[width=0.49\textwidth]{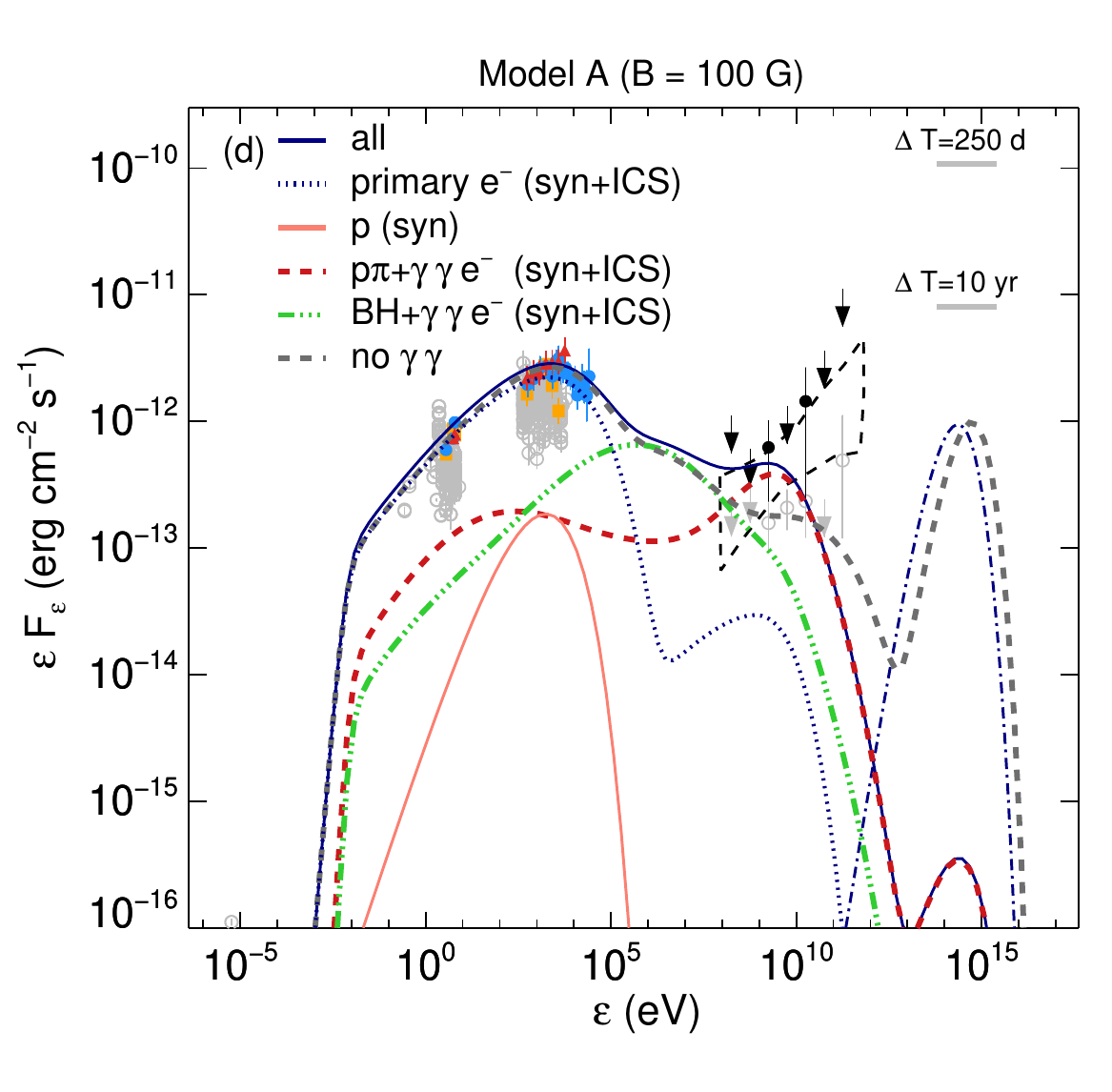}
\caption{The SED of BZB\,J0955+3551 obtained with \textit{Swift}, \textit{NICER}, and \textit{NuSTAR} data collected during 2020 January 8, 10, and 11 together with the average $\gamma$-ray spectrum of the source over a period of 250 days before the neutrino detection. Archival data are overplotted with gray open symbols. The photon spectra computed in a one-zone lepto-hadronic model (solid lines) are shown in panels (a)-(c) for three values of the magnetic field. Neutrino fluxes from the lepto-hadronic model are reported as dashed-dotted lines. The decomposition of the SED into various emission components is shown in panel (d). Image reproduced with permission from \citet{petropoulou20}, copyright by AAS}\label{0955_fig2}
\end{figure}

\begin{figure}[ht]
\centering
\includegraphics[width=0.75\textwidth]{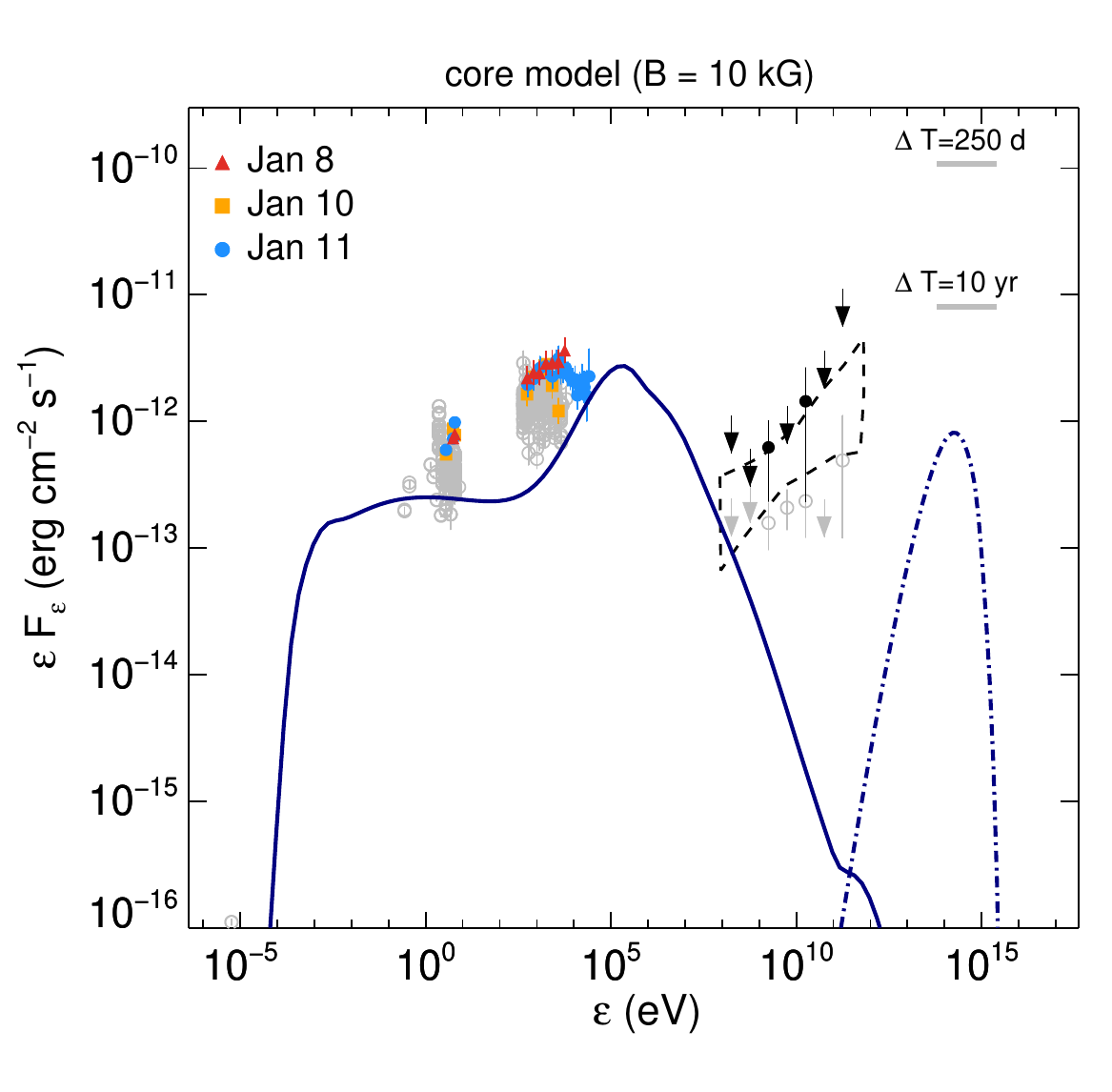}
\caption{{\it Left}: The same SED shown in Fig.~\ref{0955_fig2} at which a photon and neutrino energy spectra (solid and dashed-dotted lines, respectively) emerging from the blazar-core region is shown. Image reproduced with permission from \citet{petropoulou20}, copyright by AAS}\label{0955_fig3}
\end{figure}

In \citet{petropoulou20}, single-zone lepto-hadronic models, where the electromagnetic and high-energy neutrino emissions originate from the same region of the jet have been tested, but exploring also several alternative scenarios for neutrino production: i) a blazar-core model: neutrino production occurs in the inner jet, close to the supermassive black hole; ii) a hidden external-photon model: neutrinos are produced in the jet via interactions with photons from a weak BLR; iii) a proton-synchrotron model: high-energy protons in the jet produce $\gamma$ rays via synchrotron; iv) an intergalactic cascade model: neutrinos are produced in the intergalactic medium through interactions of a high-energy cosmic-ray beam escaping the jet. In these models, the neutrino emission is not related to the X-ray flare observed from BZB\,J0955+3551. If the association is real, future detectors like IceCube-Gen2 should be able to provide additional evidence for neutrino production in BZB\,J0955+3551 and other extreme blazars. The most optimistic models (single-zone lepto-hadronic model or multi-zone blazar core, see Figs.~\ref{0955_fig2} and \ref{0955_fig3}) predict that the long-term emission of the source would result in $\sim$ 1--3 muon neutrinos in 10 years above 100 TeV in the IceCube-Gen2 detector. The most promising scenarios for neutrino production also predict strong $\gamma$-ray attenuation above $\sim$100 GeV within the source itself.

\subsection{PKS\,0735+178}

\begin{figure}[ht]
\centering
\includegraphics[width=\textwidth]{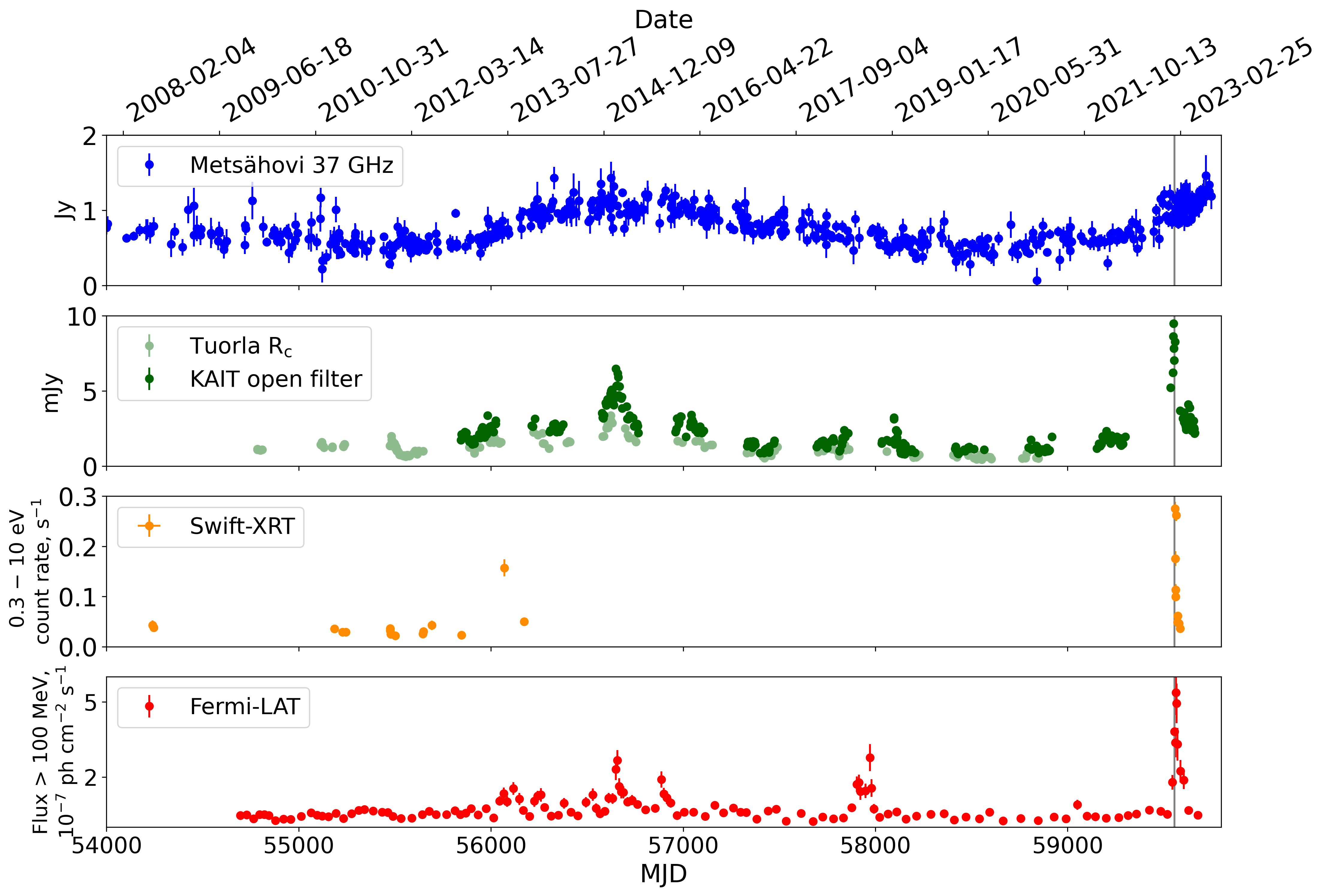}
\caption{Multi-wavelength light curves of PKS\,0735+178 collected during 2006--2023 including (from top to bottom) radio (37 GHz), optical X-ray and $\gamma$-ray observations. The grey line corresponds to the neutrino detection by IceCube. Image reproduced with permission from  \citet{omeliukh25}, copyright by the author(s)}\label{0735_fig1}
\end{figure}

In 2021 December, multiple neutrino events have been detected by different facilities: IceCube \citep{icecube21}, Baikal-GVD \citep{dzhilkibaev21}, the Bakan Underground Scintillation Telescope \citep{petkov21}, and KM3NeT \citep{filippini22}. These events were in spatial and temporal coincidence with the strongest $\gamma$-ray flare observed by the blazar PKS\,0735+178, potentially making it the first source for which multiple neutrino events have been observed. Moreover, the coincidence of a $\gamma$-ray flare and a high-energy neutrino is similar to what has been observed for TXS\,0506+056 in 2017. In addition to the $\gamma$-ray band, the source has been observed during an optical-to-X-ray flaring activity at the time of neutrino events (see Fig.~\ref{0735_fig1}). 

\begin{figure}[ht]
\centering
\includegraphics[width=0.9\textwidth]{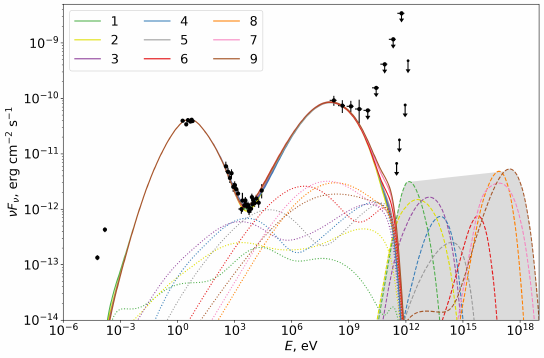}
\caption{Different lepto-hadronic models of PKS\,0735+178 that can explain photon fluxes during the neutrino arrival. The solid curves correspond to the electromagnetic emission, the dotted curves correspond to the contribution of the hadronic process in total photon fluxes, and the dashed curves correspond the predicted neutrino spectra. Image reproduced with permission from \citet{omeliukh25}, copyright by the author(s)}\label{0735_fig2}
\end{figure}

The SED of the source in different activity states has been modeled in detail in \citet{omeliukh25} to explore the physical parameters that can produce the observed neutrino events. In that work, numerical simulations of radiation processes in a one-zone model framework are presented, exploring the parameter space of leptonic and lepto-hadronic models to find best-fit solutions that explain the observed fluxes. A purely leptonic model can successfully reproduce the SEDs in quiescent, neutrino arrival, flaring, and post-flaring states using synchrotron and synchrotron self-Compton (SSC) emission components, but cannot account for neutrino emission. Adding protons in the same emitting region of the electrons does not change the SSC producing the $\gamma$-ray emission but introduces a significant hadronic contribution in the X-ray band and can explain the neutrino event (see Fig.~\ref{0735_fig2}). Under the condition of maximization of the neutrino event rates in IceCube, the lepto-hadronic model proposed a scenario in which 0.1 neutrino events were produced during the total 50-day flaring period, a higher number of neutrino events than any other previous model proposed. In addition, the maximum neutrino rate expected in a quiescent state is two orders of magnitude lower than that observed during flaring activity.
A simple one-zone lepto-hadronic model failed to explain the $\gamma$-ray fluxes in the post-flare SED, suggesting a potential need for a different contribution, for example including external photon fields e.g. from a BLR. Moreover, parameter degeneracy, in particular in hadronic parameters (i.e. the proton spectrum properties) creates great uncertainty in the neutrino emission prediction. This is related at least to the fact that the available multi-wavelength data were insufficient to unambiguously predict the neutrino spectrum. To break this degeneracy, next-generation neutrino telescopes and future X-ray and MeV polarization instruments are needed.

\section{Search for multi-flare neutrino emission from a list of $\gamma$-ray-emitting AGN}\label{sec4}

One of the properties of AGN is the extreme variability of their emission throughout the electromagnetic spectrum. Consequently, their neutrino emission may be expected to exhibit similar variability. Studies have therefore been performed to search for multi-flare neutrino emission.

In a time-integrated analysis of point-like neutrino events with energies typically above $\sim$1 TeV collected by IceCube between 2008 April 6 and 2018 July 10, a full-sky scan and a search from a catalog of 110 $\gamma$-ray emitters have been performed. A cumulative neutrino excess in the Northern Hemisphere was identified at the 3.3-$\sigma$ level using a maximum-likelihood method. The two searches (full-sky scan and catalog search) found that the hottest spot coincides with NGC 1068, with a significance of 2.9-$\sigma$. Moreover, the main contributors to the cumulative excess are NGC\,1068, TXS\,0506+056, GB6 J1542+6129 and PKS\,1424+240 \citep{aartsen20}. 

Although time-integrated searches are highly sensitive, they also highlight the need for time-dependent analysis in which the suppression of the time-constant background of atmospheric neutrinos is favoured and potential high-activity periods are particularly suitable for neutrino production in jets. A separate time-dependent neutrino emission search on the same data and source catalog by the IceCube Collaboration found M87 as the most significant time-dependent source of the catalog at 1.7-$\sigma$ significance level post-trial and TXS\,0506+056 the only source with two flares observed. The significance of the detection of TXS\,0506+056 is lower than the value reported in \citet{aartsen18c}, and is mainly related to the different selection of events used in the analysis \citep[see Appendix E in][]{abbasi21}. 
In addition, based on a binomial population test, a cumulative time-dependent neutrino excess has been identified in the Northern Hemispehere at the 3.0-$\sigma$ level, associated with four sources: M87, TXS\,0506+056, GB6 J1542+6129 and NGC\,1068 \citep{abbasi21}. Three of the four sources are in common with the time-integrated analysis, with only M87 showing a strong time-dependent behaviour. 

These two analysis methods, time-dependent and time-integrated, confirm that only a small population of neutrino-emitting AGN are identified in the Northern Hemisphere, at least at the current sensitivity level of the IceCube detector, and they might show a time dependent trend. In addition, no significant excess has been identified in the Southern Hemisphere, consistent with the lower sensitivity related to the larger background of atmospheric muons in that hemisphere. 

Among the sources associated with a neutrino excess in the time-integrated or time-dependent analysis, GB6\,J1542+6129 is claimed to share some properties with TXS\,0506+056 \citep{padovani22}, and, although there is no consensus on its redshift, a multi-wavelength analysis of this blazar suggests a disturbance-driven, multi-zone scenario where high-energy neutrinos are produced in a compact, photon-rich region near the SMBH with a delayed temporal relationship between neutrino and electromagnetic flares \citep{kun26}. In the following subsections, we briefly discuss some interesting properties of PKS\,1424+240 and M87.

\subsection{PKS\,1424+240}

\begin{figure}[ht]
\centering
\includegraphics[width=0.8\textwidth]{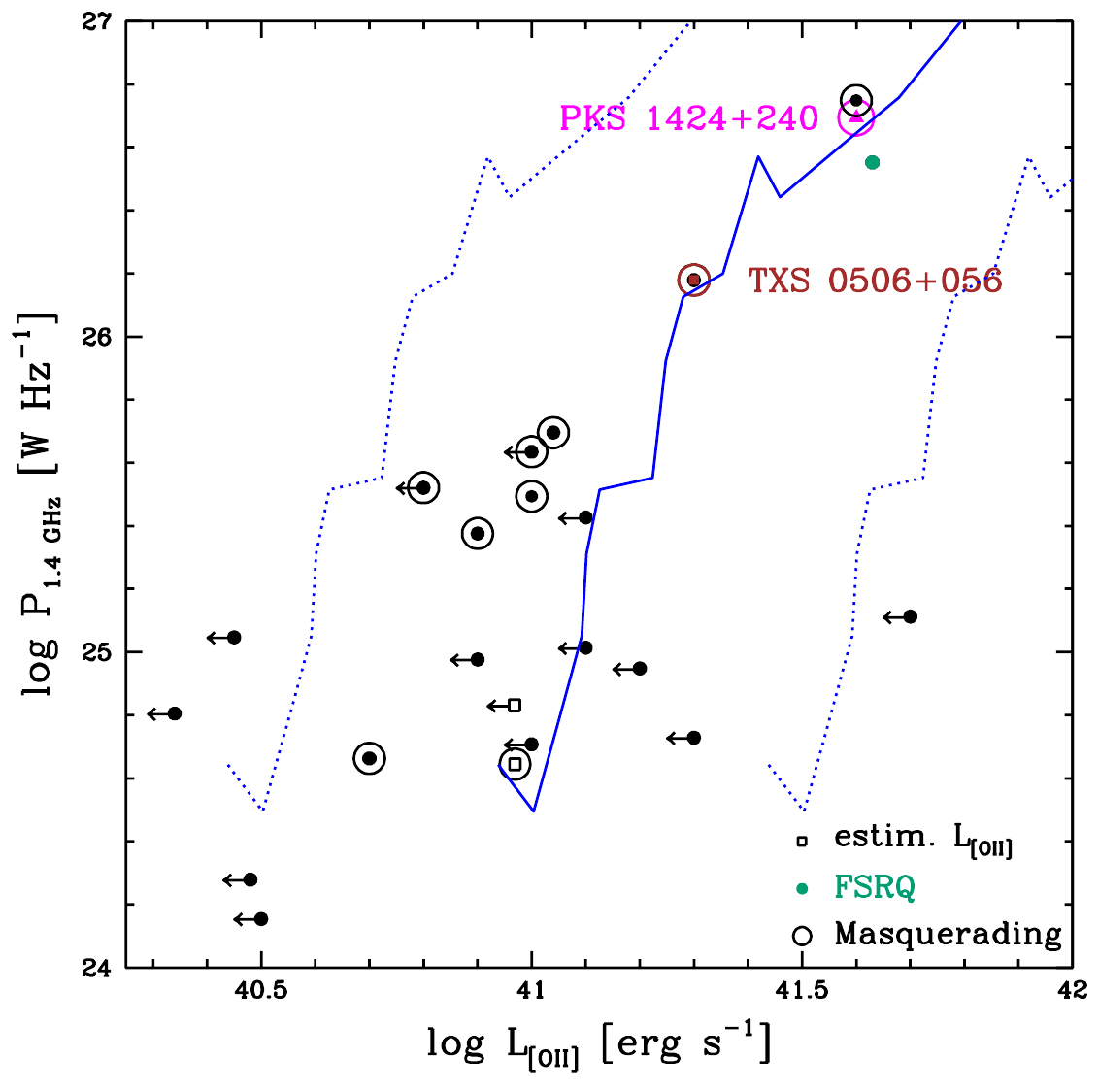}
\caption{Plot of the P$_{1.4\,\mathrm{GHz}}$ versus $L_\mathrm{[OII]}$ of a sample of neutrino candidate sources (black filled circles), with masquerading sources reported as larger empty circles and PKS\,1424+420 and TXS\,0506+056 highlighted. Sources for which $L_\mathrm{[OII]}$ has been estimated from $L_\mathrm{[OIII]}$ are denoted by black empty squares. The green filled circle denotes the FSRQ, while the solid blue line is the locus of jetted quasars, with the two dotted lines indicating a spread of 0.5 dex. Black empty squares are reported when the $L_\mathrm{[OII]}$ is estimated by the $L_\mathrm{[OIII]}$. Image reproduced with permission from \citet{padovani22}, copyright by the author(s)}\label{masquerade}
\end{figure}

PKS\,1424+240 has been claimed to be a masquerade BL Lac, similar to TXS\,0506+056. Other properties seem to be shared between these two potential neutrino emitting AGN, i.e. the overall SED, high jet power, pc scale properties \citep{padovani22}. For example, in the radio power vs emission line power diagram (see Fig.~\ref{masquerade}) defined in \citet{kalfountzou12}, PKS\,1424+240 and TXS\,0506+056 seem to lie in the locus of the jetted/radio-loud quasars. 
Those two sources can be classified as high-excitation galaxies, providing a rich radiation field external to the jet, and thus more target photons for the protons involved in the production of neutrinos. These similarities may suggest that neutrino-emitting jetted AGN can be part of a subclass with specific properties.

The analysis of VLBA observations at 15 GHz suggests that the angle of view for this blazar is $<$ 0.6$^{\circ}$, with the source observed inside the jet cone. This close alignment allows detection in polarimetric observations of a strong, net toroidal magnetic field component, indicating a powerful, current-carrying jet flowing towards us. This configuration can provide the necessary environment to accelerate protons to the high energies required for neutrino production in this source \citep{kovalev25}.

\subsection{M87}

Based on its $\gamma$-ray properties, M87 has been considered a misaligned BL Lac object, in particular a misaligned high-energy-peaked BL Lac (HBL) \citep{reimer04}. Moreover, a persistent central ridge structure, a spine flow in the M87 jet, in addition to a limb-brightening structure, have been observed in M87 with high-resolution radio images (e.g., \citet{hada24} and the references therein). A similar spine-layer structure has also been observed in TXS\,0506+056 \citep{ros20}, where the core jet expands in size with apparent superluminal velocity. This can be related to a deceleration of the jet due to proton loading from a spine-sheath structure \citep[e.g.,][]{ghisellini05,tavecchio08}, and is suggested as a possible explanation of the higher neutrino flux with respect to the $\gamma$-ray flux observed by the MAGIC telescopes \citep{ansoldi18}. An Advection-Dominated Accretion Flow (ADAF) is considered the favoured theoretical framework to explain the extremely low luminosity and high-power jet of its SMBH. In this context, the high ion temperatures supply a reservoir of relativistic protons. However, to increase the efficiency of neutrino production, high values of the proton power are needed. Alternatively, target photon fields external to the jet can be efficient. In this context, a spine-layer scenario, like the structured jet, is a very intriguing scenario, in which the density of photons originating from the slower layer is highly boosted in the comoving frame of the faster spine, enhancing the targets for proton-$\gamma$ interactions and thus neutrino emission without requiring an extreme proton power.

\section{Correlation between high-energy neutrinos and $\gamma$-ray-emitting AGN}\label{gamma}

\subsection{Correlation between astrophysical neutrino and \textit{Fermi}-LAT AGN}

The connection between high-energy neutrinos and $\gamma$ rays has long been investigated in multi-messenger astronomy, since both signals can originate from the same extreme physical processes. In fact, neutrinos and $\gamma$ rays are produced simultaneously when protons accelerated to relativistic speeds interact with matter or radiation fields. After the confirmed connection between the blazar TXS\,0506+056 and IceCube-170922A, many studies have focused on the connection of neutrinos with individual blazars, investigating the multi-wavelength behaviour of candidate neutrino emitting AGN \citep[e.g.][]{garrappa19,franckowiak20}. 

\begin{figure}[ht]
\centering
  \includegraphics[width=0.32\linewidth]{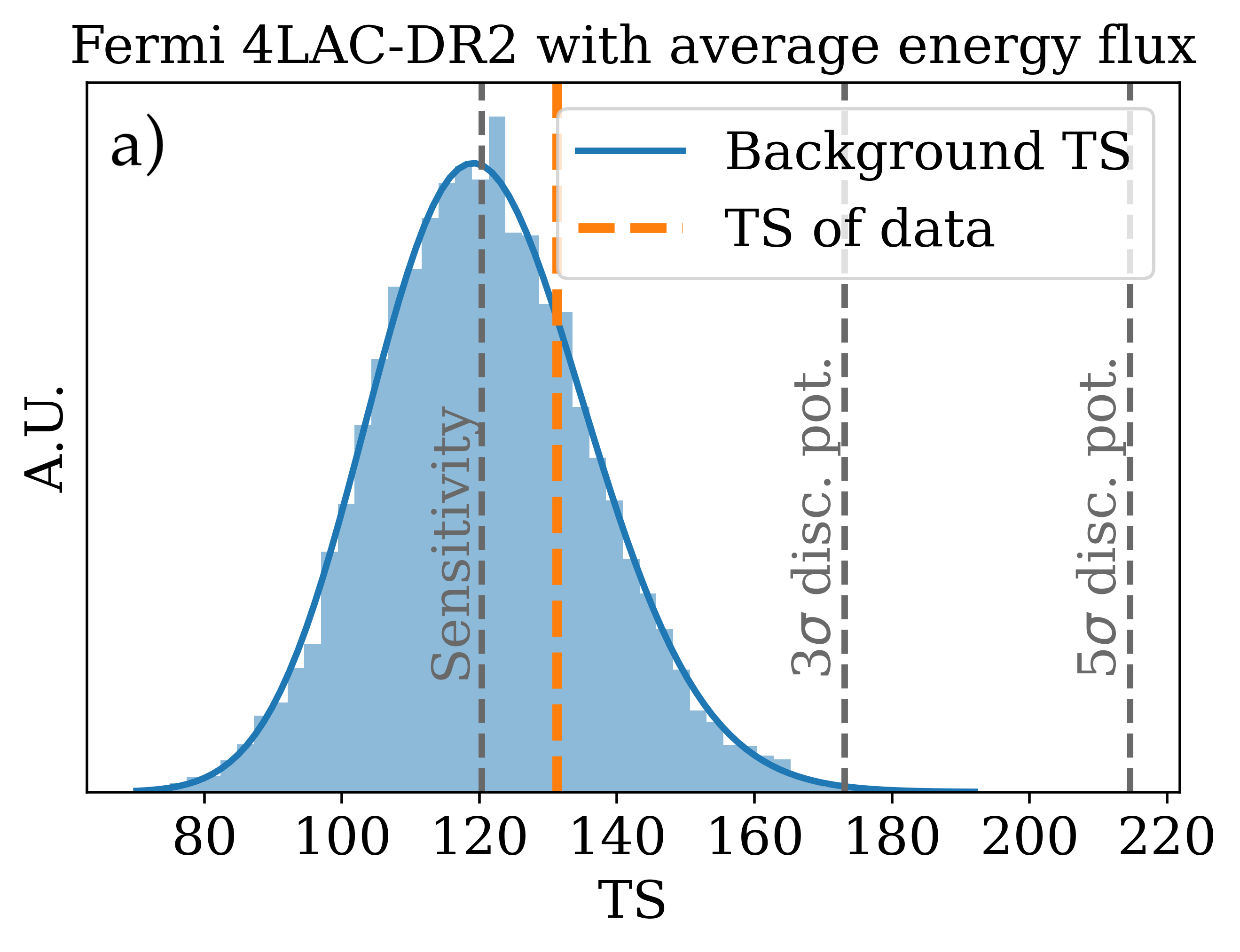}
  \includegraphics[width=0.32\linewidth]{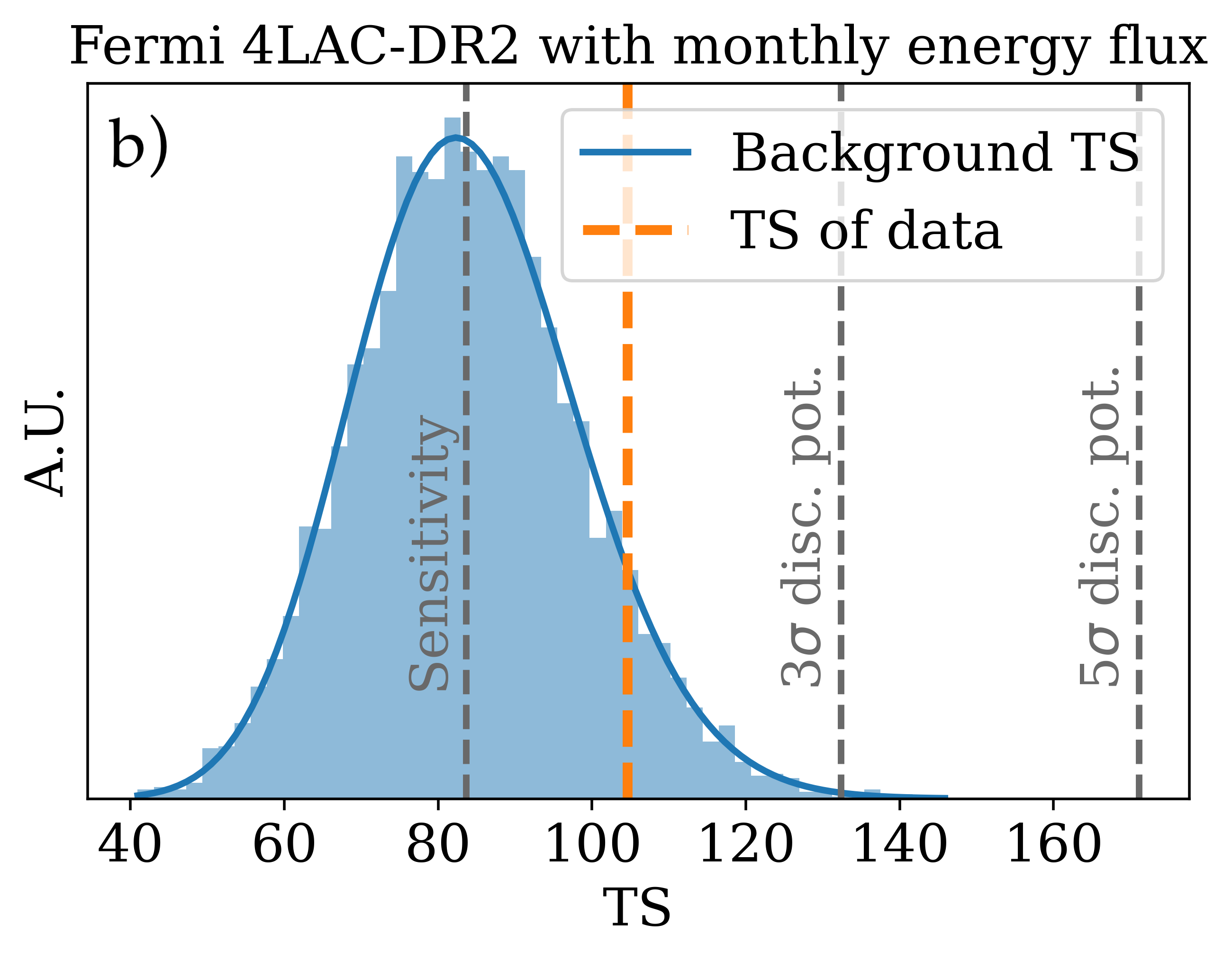}
  \includegraphics[width=0.32\linewidth]{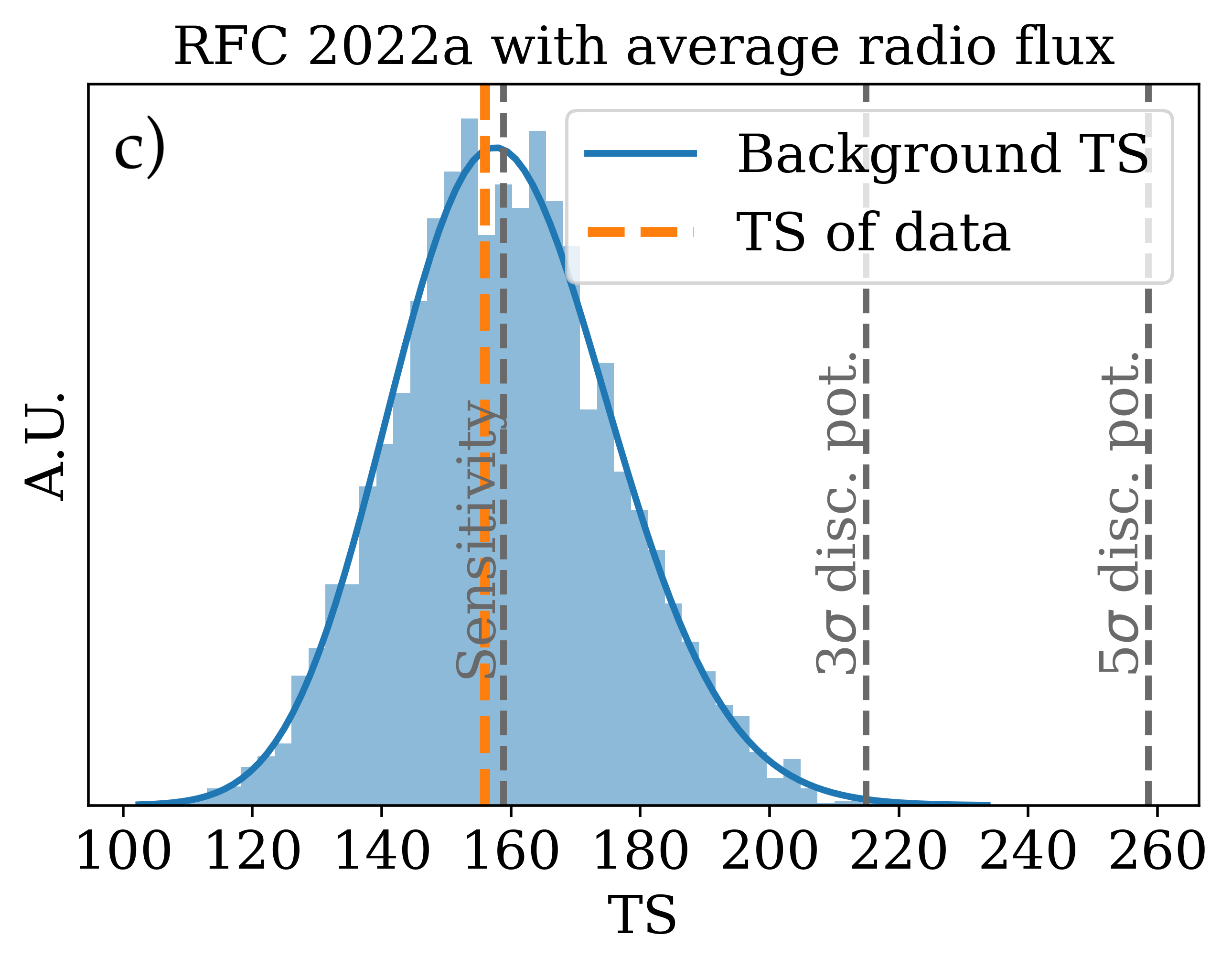}
\caption{The blue histogram shows the distribution of the TS values for N = 5000 background maps, fitted to a gamma distribution represented as a blue curve. The orange dotted line represents the observed TS of the data, while the gray lines represent the TS needed to reject the background hypothesis at the 3-$\sigma$ or 5$\sigma$ level for the 4LAC-DR2 sample with the average energy flux weights (left panel) and the monthly energy flux weights (middle panel), respectively, and for the RFC sample (right panel). Image reprodcued iwth permission from \citet{abbasi23b}, copyright by the author(s)}\label{gamma_fig1}
\end{figure}

More recently, there has been an increasing number of studies focused on the connection between high-energy neutrinos and blazar populations at different wavelengths, starting from the $\gamma$-ray band \citep[e.g.][]{krauss18, giommi20}. However, no statistical correlation has been found between $\gamma$-ray blazars and high-energy neutrinos detected by IceCube so far \citep{aartsen17, abbasi23}. In particular, in \citet{abbasi23} the correlation between the first catalog of alerts that IceCube has published \citep[IceCat-1;][]{abbasi23b} and both 2089 blazars from the Fourth Fermi-LAT catalog data-release 2 \citep[4LAC-DR2][]{ajello20,lott20} and 3413 AGN from the Radio Fundamental Catalog \citep[RFC;][]{petrov25} have been studied using a stacking analysis method to calculate the statistical significance of the overall correlation. There was no significant correlation between the 4LAC blazars weighted by the average 10-year energy flux and the neutrino alerts, with a p-value of 0.248 (see Fig. \ref{gamma_fig1}, left panel). Among the 10 pairs of blazar-neutrino with the highest Test Statistic (TS), the two strongest correlations have been found for TXS\,0506+056 and PKS\,1502+102 with IC170922A and IC190730A, respectively. The third and fourth correlations have been for 3C 454.3 and Mrk 421, two radio bright sources, but the area of the error region of the associated neutrino is very large (i.e. 90\% contour of 26.80 deg$^{2}$ and 56.07 deg$^{2}$, respectively). A similar result has been obtained when the selected blazars are weighted by the energy flux at the neutrino arrival time, with a p-value of 0.08 (see Fig. \ref{gamma_fig1}, middle panel), except 3C~454.3 being not among the pairs with the highest TS because the source was in a quiescent state at the time of neutrino detection. 

A correlation between neutrino events using 7 years of IceCube data and the position of blazars from a sample of objects in the 5BZCat in the Southern Hemisphere has been claimed by \citet{buson22} but, as discussed in \citet{bellenghi23}, this correlation is strongly disfavoured. In particular, extending the search to the Northern Hemisphere, where IceCube is most sensitive to astrophysical signals, no significant correlation between 5BZCAT blazars and neutrino is found in \citet{bellenghi23}. Broadening the observational IceCube period to 10 years of data, the correlation disappears in both the Southern and Northern Hemispheres, suggesting that the potential correlation reported in \citet{buson22} can be a statistical fluctuation and also related to the fact that the sample used for blazars is not derived from a well-defined, flux-limited catalog. 


\subsection{Correlation between IceCube neutrino events and the \textit{Fermi} unresolved $\gamma$-ray sky}

Unlike the previous approach, a cross-correlation analysis of the neutrino events not with the $\gamma$-ray-detected AGN but with potential faint sources not resolved individually by \textit{Fermi}-LAT has been carried out in \citet{negro23}. In particular, the population of sources contributing to the unresolved $\gamma$-ray background (UGRB) has been considered for the analysis. Three main assumptions are used: i) no intrinsic cut-off of the spectrum above the \textit{Fermi}-LAT energies; ii) no hardening of the spectrum above the \textit{Fermi}-LAT energies; iii) blazars being the dominant population for the UGRB. Detailed analyses of the 2D spatial cross-correlation between IceCube muon-neutrino events and the \textit{Fermi}-LAT UGRB intensity maps did not identify a significant signal. If the $\gamma$-ray emission of the UGRB is produced by neutral pions from the p--p or p--$\gamma$ interaction, up to 60\% or 30\%, respectively, of the specific population of sources that dominate the UGRB fluctuations may contribute to total neutrino events, although their overall impact on the total astrophysical flux remains small. While the UGRB itself is known to be dominated by unresolved blazars, the results reported in \citet{negro23} suggest that these blazars are not the primary drivers of the observed diffuse neutrino flux detected by IceCube, and only a contribution of $<$ 1\% of the observed astrophysical neutrino flux at 100 TeV could be explained by the decay of pions created in $\gamma$-ray dim unresolved blazars in the \textit{Fermi}-LAT UGRB.

\subsection{Cross-correlation between IceCube neutrino events and VHE AGN observed by IACT}

Simultaneous multi-frequency observations can be important in characterizing the neutrino sources. In this context, Very High Energy (VHE; $E$ $>$ 100 GeV) $\gamma$ rays are produced together with neutrinos, when charged and neutral pions, produced in hadronic and photohadronic interactions, decay into neutrinos and $\gamma$ rays. 

\begin{figure}[ht]
\centering
\includegraphics[width=0.8\textwidth]{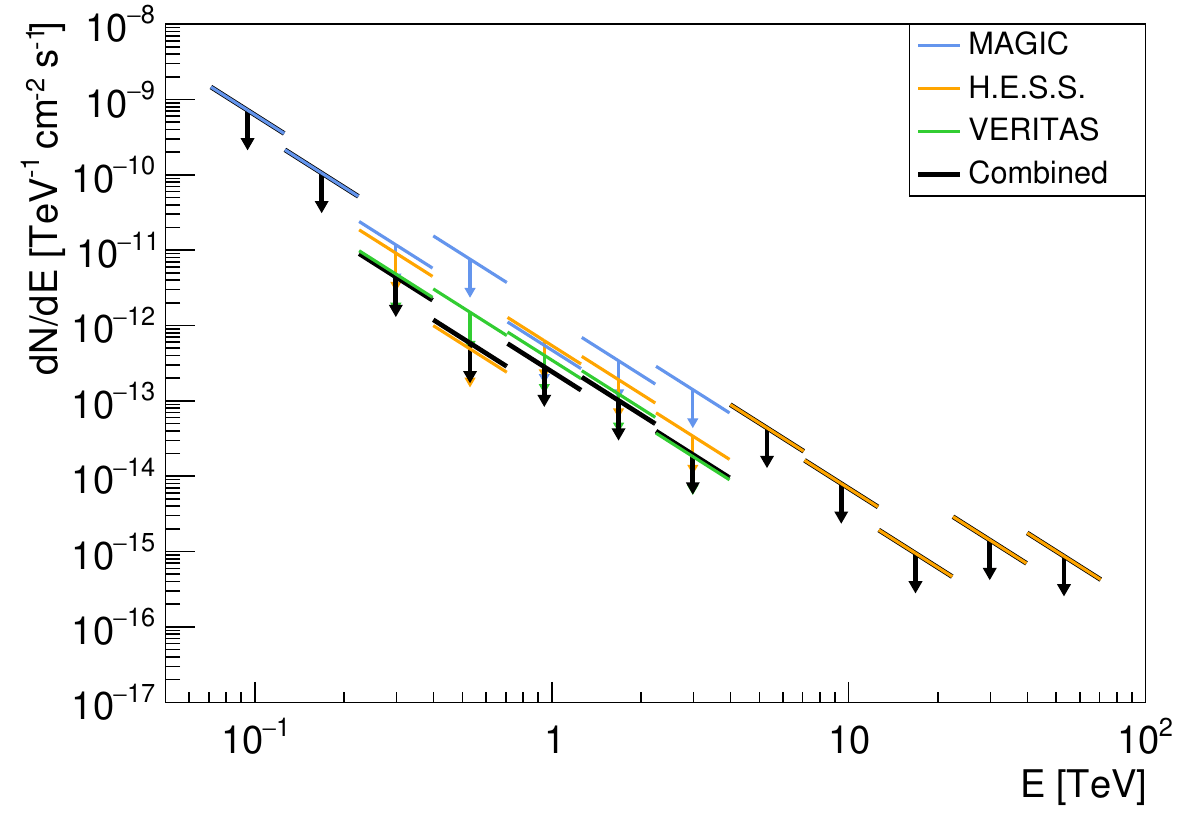}
\caption{Combined differential-flux upper limits at 95\% confidence level for 4FGL\,J0658.6+0636 associated with IC-201114A using IACT. Image reproduced with permission from \citet{abhir25}, copyright by the author(s)}\label{IACT_fig1}
\end{figure}

\begin{figure}[ht]
\centering
\includegraphics[width=0.8\textwidth]{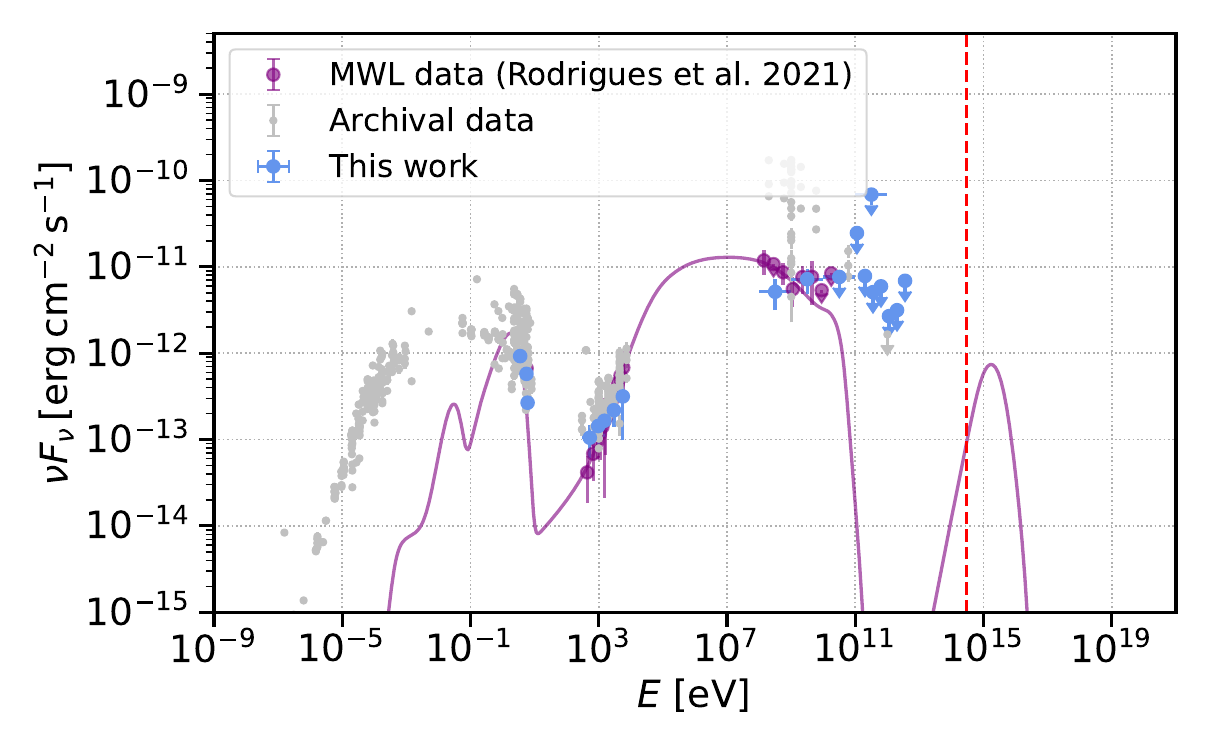}
\caption{SED modelling of the blazar PKS\,1502+106, potentially associated with the single high-energy neutrino alert IC-190730A, compared with IACT upper limits and simultaneous multi-frequency and archival data. Image reproduced with permission from \citet{abhir25}, copyright by the author(s)}\label{IACT_fig2}
\end{figure}

For this reason, follow-up observations of neutrino events with current imaging atmospheric Cherenkov Telescopes (IACT) can provide an important confirmation of the neutrino-EM signal and new insights about the acceleration mechanisms in these objects, as demonstrated by the case of TXS\,0506+056 and the connection with IceCube-170922A. Four currently operating IACT systems, FACT \citep{biland14}, H.E.S.S. \citep{aharonian06}, MAGIC \citep{aleksic12}, and VERITAS \citep{holder06} have conducted a follow-up program of the most promising neutrino events in cooperation with IceCube. Two different observing strategies have been carried out, depending on the neutrino trigger, i.e. neutrino event clusters or single high-energy neutrino events. In the former case, the alert consists of a cluster of candidate neutrino events with $E > 1\, \mathrm{TeV}$ detected by IceCube from a possible source within a time window; these alerts are distributed by IceCube to IACTs under a memorandum of understanding within the framework of the Gamma-ray follow-up (GFU) program since 2012. In the latter case, the alerts refer to single high-energy neutrino events (E $>$ 60 TeV), likely of astrophysical origin, publicly distributed. ToO observations with the IACTs have been triggered in the case of a promising AGN candidate identified by multi-wavelength observations by \textit{Fermi}-LAT or other electromagnetic observatories. 

The results of the analysis of follow-up observations in the period between 2017 September and 2021 January have been reported in \citet{abhir25}. In particular, the IACTs have performed follow-up observations of six GFU-cluster alerts and one all-sky cluster alert during 2019--2020, and follow-up observations of eleven single high-energy neutrino events during 2017 September-2021 January. Potential counterpart sources are identified within the uncertainty region for six events. However, no associations have been found between VHE $\gamma$-ray sources and neutrino events, but combined upper limits on the VHE flux have been obtained by a joint analysis of data from different IACT (see e.g. Fig. \ref{IACT_fig1}). On the other hand, the TeV observations together with X-ray observations available may be useful for constraining the maximum contribution from photo-hadronic interactions (see e.g. Fig. \ref{IACT_fig2}). 

\section{Correlations between astrophysical neutrino and radio bright AGN}\label{radio}

The mechanisms of neutrino production are still debated, and in the case of photo-hadronic interactions in AGN, a correlation of neutrinos is expected not only with $\gamma$ rays but also with photons emitted in radio. Although the radio emission is synchrotron self-absorbed close to the central engine, and thus what we observe is the emission originated several parsecs downstream from the jet, flux densities in radio can increase before a strong flaring activity at higher energy bands begins. In this context, the radio band can be a good proxy for jet activity.

A search for the association between individual high-energy neutrinos and long-term radio flares from blazars observed by the Owens Valley Radio Observatory (OVRO) and Mets{\"a}hovi Radio Observatory resulted in  post-trial spatial correlations of $\sim$2-$\sigma$ between radio bright AGN and astrophysical high-energy neutrinos \citep{hovatta21}. However, in some cases, they find a connection between the largest flares in radio and neutrino arrival times that is unlikely to be a random coincidence. This is in agreement with the findings reported in \citet{plavin25}, for which candidate neutrino-emitting AGN have strong jet beaming compared to other radio-bright sources and the complete uniformly selected MOJAVE sample.

\citet{kouch24} performed a spatio-temporal association analysis between 1157 blazars in the Candidate Gamma-Ray Blazar Survey (CGRABS) catalog \citep{healey08} with OVRO radio light curves and 1061 blazars in the CGRaBS catalog with Catalina Real-Time Transient Survey (CRTS), Asteroid Terrestrial-impact Last Alert System (ATLAS), and Zwicky Transient Facility (ZTF). Similarly to \citet{hovatta21}, even using two more years of OVRO data, a post-trial 2.17-$\sigma$ spatio-temporal correlation between the emission of CGRaBS blazars and neutrinos has been obtained. It is worth noticing that this correlation value has been obtained  adding an upper limit of the IceCube systematic errors of $1^{\circ}$ in quadrature to the error bars. When the IceCube error regions are taken as their published values, no significant correlation is obtained in \citet{kouch24}.  

Instead, a tentative correlation between neutrinos and radio blazars was claimed by \citet{plavin20, plavin21}\footnote{An updated analysis adding 14 new IceCube events provided compatible results \citep{plavin23}.}. However, these results are not confirmed by the analysis presented in \citet{abbasi23}. In particular, no statistically significant correlation has been observed between RFC sources weighted by the average VLBI flux at 8.6 GHz and neutrino alerts, with a p-value of 0.57 (see Fig.~\ref{gamma_fig1}, right panel). This analysis of the correlation between bright radio AGN and neutrino alerts is similar to what was performed in \citet{plavin20}, based on the same AGN catalog. The discrepancies arise primarily from three factors: i) neutrino angular uncertainties are scaled up in \citet{plavin20}, artificially enlarging the error regions, while the analysis in \citet{abbasi23} avoids scaling errors, treating published alert coordinates as rigorously reported; ii) the correlation search in \citet{abbasi23} integrated a more sophisticated description of the neutrino spatial probability density function; iii) a larger number of events has been included in \citet{abbasi23}. In particular, factor i) indicates that the AGN that contributes to the TS can come from outside the original contours and thus away from the best-fit position. 

\noindent A similar analysis, searching for the correlation between RFC sources and neutrino events collected by the ANTARES neutrino telescope over 13 years of operation, resulted in a significance level of 2.2-$\sigma$ \citep{albert24}. The lack of significant correlation between radio bright AGN and high-energy neutrinos suggests that $<1\%$ of the studied AGN can emit neutrinos related to those IceCube alerts. Using a stacking analysis with a time-dependent approach between 15 GHz radio observations of AGN in the MOJAVE XV catalog and 10 years of neutrino data, no significant correlation was found in \citet{abbasi24}. 

\section{Future prospects}\label{sec7}

\subsection{Open questions}

The detection of an ultra-high-energy neutrino by KM3NeT has opened a new window in neutrino astrophysics, but also new questions and challenges. First of all, this event challenges our understanding of the cosmic neutrino spectrum, which, based on previous observations with IceCube, seems to show a steep decline at the highest energies \citep{abbasi22}. A clear connection between this event and an extragalactic source, in particular AGN, has not been established, although three candidates have been identified: PKS\,0605$-$085, which experienced a $\gamma$-ray flaring activity detected by \textit{Fermi}-LAT and peaked around 181 days before the neutrino event; PMN\,J0606$-$0724, which exhibited a strong radio flare peaking just 5 days prior to the neutrino's arrival; MRC\,0614$-$083, which showed a multi-year, long-term increase in its X-ray light curve leading up to the neutrino event \citep{adriani26}. The possibility that this neutrino is cosmogenic, i.e., produced by UHECR interacting with cosmic background radiation has been proposed \citep{adriani25}. More generally, such extreme events can be the indication of a new class of extreme cosmic accelerators with respect to the other high-energy neutrinos detected so far. Therefore the detection of this record-breaking, ultra-high-energy neutrino in 2023 challenges existing models of cosmic neutrino spectra, requiring more KM3NeT data, potentially with electromagnetic counterparts, to pinpoint origins and refine our understanding of these events.

The discovery of a diffuse astrophysical neutrino flux in the 10 TeV--10 PeV range by IceCube marked a fundamental discovery, with the statistical significance of the detection increasing as the statistics increase, and different analysis channels have been used for the analysis, but the identification of the sources contributing to this flux remains a fundamental open question. The primary challenge lies in the fact that most detected neutrinos cannot be traced back to specific point sources with certainty, while a diverse range of astrophysical objects has the potential to serve as neutrino emitters. Due to the typical error region of neutrino events, chance spatial coincidences are not conclusive, and additional evidence for a robust neutrino-source association is needed. Both population studies and multi-wavelength variability studies are important for characterizing neutrino sources. In this context, multi-wavelength and time-domain astronomical observations are increasingly important to improve the identification of possible neutrino-emitting AGN. 

Blazars constitute the main source of the extragalactic diffuse $\gamma$-ray background \citep{ajello15}, thus if these $\gamma$-ray photons are produced only by hadronic interactions, blazars should provide a significant contribution to the diffuse neutrino background. This indicates the presence of a class of high-energy neutrino sources opaque for $\gamma$ rays, suggesting that p-$\gamma$ is the main production channel for these neutrinos. One debated issue related to neutrino-emitting blazars is the importance of the Doppler boosting. While Doppler boosting can significantly increase the observed neutrino signal, it does not necessarily change the intrinsic physical efficiency of neutrino production inside the source. Open questions are whether neutrinos are produced isotropically in the jet frame, how strongly their production regions are boosted by relativistic motion, and whether they perfectly correlate with electromagnetic emission. Observationally, researchers have found a possible connection between high-energy neutrino events and blazars that feature extremely bright, parsec-scale radio emission. Studies analyzing data from the MOJAVE program indicate that neutrino-candidate blazars demonstrate extreme jet beaming, with smaller viewing angles and higher Lorentz factors (e.g., \citealt{plavin25}). However, directional uncertainties of IceCube events make it difficult to uniquely pair high-energy neutrinos with specific radio-bright blazars and their exact correlation is still debated.

To better characterize the neutrino emitting sources will be important to refine the measurements of the astrophysical neutrino observables (i.e. distribution of arrival directions and arrival times, energy spectrum, and neutrino flavour composition), and for doing this, not only more data from the current neutrino facilities but observations from future facilities will be fundamental, even more for extending the study of these astrophysical neutrino observables beyond 10 PeV. Three important characteristics of neutrino facilities are timing resolution, pointing resolution, and sky coverage. An improvement in these characteristics will allow us to probe the most energetic sources in the Universe, indirectly with the diffuse flux, and directly with the discovery of point sources associated with neutrino emission in temporal or spatial coincidence with cosmic rays and electromagnetic emission. Moreover, observations of the neutrino characteristics are significantly more effective if made in synergy with other messengers, such as photons, cosmic rays and gravitational waves. In Sect.~\ref{facilities} current and future facilities are briefly described, while the landscape from radio to $\gamma$ rays of the observatories useful for searching for AGN/neutrino connection in the next decade and so is discussed in Sect.~\ref{MWL}.

\subsection{Current and future neutrino facilities}\label{facilities}

Current neutrino detectors are based on the water Cherenkov technique \citep{reines60,greisen60,markov60}, in which glacial ice, a deep sea or a deep lake can be used as a transparent medium. When a neutrino interacts with an atomic nucleus in this medium, it produces secondary charged particles that travel faster than the speed of light in that medium, creating a faint blue glow of Cherenkov radiation. Only one neutrino observatory is fully operational in mid-2026: IceCube. In addition,  ANTARES \citep{albert21} was operative until February 2022, which marked the end of its data-taking phase. ANTARES was located in the Mediterranean Sea near the coast of Marseille (France),  reaching its final configuration in 2008 with 885 photo-multiplier tubes (PMT) for a total volume of 0.01 km$^{3}$. 

\begin{figure}[ht]
\centering
\includegraphics[width=\textwidth]{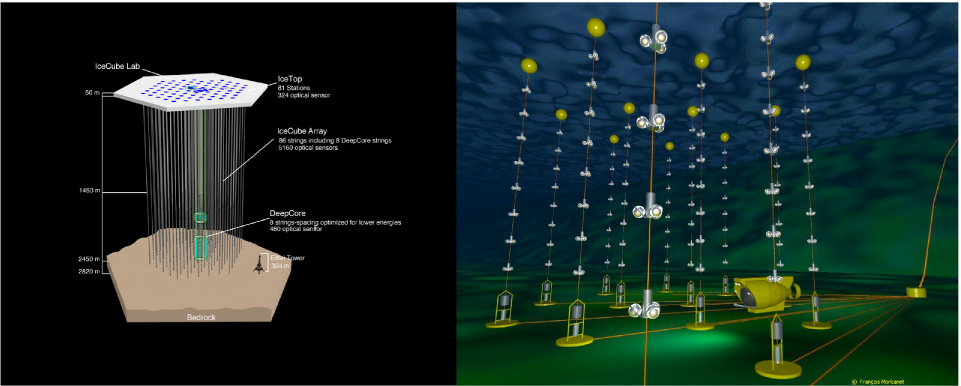}
\caption{The current generation of high-energy neutrino telescopes: IceCube (left) and ANTARES (right)}\label{IceCubeANTARES}
\end{figure}

The IceCube Neutrino Observatory is a neutrino detector occupying an instrumented volume of 1 km$^{3}$ (Fig.~\ref{IceCubeANTARES}) within the Antarctic ice sheet at the Amundsen-Scott South Pole Station optimized for the detection of high-energy neutrinos above 100 GeV \citep{ahrens04}. The detector was completed in 2010 and consists of 5160 photomultipliers tubes distributed in an array of 86 vertical strings, nominally spaced 125 m apart, at a depth of $\sim$1.5 to 2.5\,km on the ice. In particular, IceCube includes IceTop, an air-shower array on the ice surface for studying the cosmic-rays, and DeepCore, a deeper and denser array optimized for studying high-energy neutrinos. The next generation IceCube detector, IceCube-Gen 2 \citep{aartsen21}, will increase the volume of the detector and its sensitivity to high-energy neutrinos.

New improvements are expected with the completion of the KM3NeT neutrino telescope \citep{adrian16} in the Mediterranean Sea and the extensions of the Gigaton Volume Detector (GVD) in the Lake Baikal \citep[Baikal-GVD;][]{ageron11}. KM3NeT is made up of two components: the Oscillation Research with Cosmic rays in the Abysses (ORCA), dedicated to studying neutrino properties, and the Astronomy Research with Cosmic rays in the Abysses (ARCA), optimized for studying high-energy neutrinos. In the final configuration, ARCA will be made up of two blocks of 115 strings (i.e. detector units) with 18 optical modules per string for a total collective area of $\sim 1\, \mathrm{km}^{3}$. ORCA will comprise a single block of 115 detection units, with a total of 345 detection units for the full observatory. In mid-2026, KM3NeT has around 84 detection units installed and running, forming the initial core of the ARCA and ORCA detectors. 

\noindent The Baikal-GVD is the largest operating neutrino telescope in the Northern Hemisphere and is in the final phase of construction toward a 1-cubic-kilometer volume. As of late 2025, the array consists of 14 sub-arrays (clusters), which include 117 strings holding 4212 optical modules submerged at depths between 750 and 1275 m in Lake Baikal covering a volume of 0.7 km$^{3}$ \citep{safronov25}. The final configuration of Baikal-GVD will be obtained by 2028 reaching its full effective volume of 1.0 km$^{3}$. In addition, other neutrino telescopes have been proposed, such as P-ONE \citep{resconi22}, TRIDENT \citep{ye22}, RNO-G \citep{aguilar21}, and GRAND \citep{fang17}. 

KMNeT, P-ONE, and Baikal-GVD will improve our sensitivity to TeV-PeV neutrinos in the Southern Sky. Finally, the next generation of the IceCube detector, IceCube-Gen2, will increase the volume, significantly improving the sensitivity to high-energy neutrinos, from the TeV scale to the EeV scale. All these facilities will be able to identify more astrophysical neutrino events, improve the angular localization, and, thanks to the new facilities in the Northern Hemisphere, be able to complemented IceCube for the study of the Galactic Plane. 

\subsection{The multi-wavelength landscape searching for the AGN/neutrino connection in the next decade}\label{MWL}

Since 2016, IceCube has automatically reported the detection and localization of single high-energy neutrinos in near real-time \citep[e.g.][]{aartsen17b}, enabling multi-wavelength follow-up observations to characterize these events. Neutrino follow-up programs across the electromagnetic spectrum have identified candidate extragalactic sources, primarily AGN in addition to Tidal Disruption Events (TDE). In particular, to capture transient behaviour of candidate AGN neutrino emitters, it is important to have facilities able to send and respond to real-time alerts, thus continuous operation and a complete cover of the sky are fundamental for detecting transient activities in the electromagnetic spectrum and improving overall flux sensitivity.

A geographical distribution of the ground-based observatories at different wavelengths in latitude will be fundamental for full-sky coverage across the Northern and Southern hemispheres to improve the chance of having a prompt follow-up observation of a neutrino event and a continuous observation after the event. In this context, wide-field electromagnetic facilities or pointed telescopes that can scan the same part of the sky with high cadence. In addition, prompt target of opportunity observations and long-term monitoring can be very important for giving a look at the electromagnetic spectrum at the time of the neutrino arrival or near to, and for better characterizing the source activity over a long time-scale.

The upcoming Cherenkov Telescope Array Observatory (CTAO) explicitly mentions neutrino follow-up observations as part of the Key Science Projects to be conducted in the early operation \citep{CTAO19}, and thanks to telescopes in both the hemispheres and improved performances in terms of sensitivity and angular resolution with respect to current IACTs promises to make a significant step forward in the follow-up of high-energy neutrino events at VHE. CTAO observations will be complemented with the monitoring provided by wide-field instruments already in operation, such as the High-Altitude Water Cherenkov (HAWC) \citep{abeysekara17} and the Large High Altitude Air Shower Observatory (LHAASO) \citep{aharonian21}, in the Northern Hemisphere. In the Southern hemisphere, there are currently no wide-field instruments operating at TeV energies, but in the future the Southern Wide-field Gamma-ray Observatory (SWGO) \citep{hinton22} will fill this gap. In the GeV band, \textit{Fermi}-LAT is scanning the $\gamma$-ray sky since 2008 \citep{atwood09}. A potential problem will arise in the case that \textit{Fermi} stops operation, considering that there are no future GeV missions in an advanced stage of conception.

The X-ray band, in general, is important for tracing the energetic processes associated with neutrino production. In blazars, X-rays trace non-thermal jet activity, with flares potentially marking periods of enhanced neutrino emission and specific features. In Seyfert galaxies, X-rays originate from the hot corona above the accretion disk, a potential site of proton acceleration. In particular, the hard X-ray to MeV energy range can be important for investigating neutrino-emitting AGN. In hard X-rays \textit{NuSTAR} \citep{harrison13} is the most sensitive telescope in operation. There is a gap between the GeV regime and the hard X-rays, with no current mission expected to cover the MeV regime. This gap will be closed by the upcoming Compton Spectrometer and Imager (COSI) mission, a NASA Small Explorer satellite mission in development with a planned launch in 2027 \citep{tomsick23}, which operates in an energy range of 0.2 to 5 MeV, and possibly the All-sky Medium Energy Gamma-ray Observatory eXplorer (AMEGO-X), with an energy coverage from 25 keV to 1 GeV \citep{caputo22}. Thanks to the fast repointing capability, in soft X-rays \textit{Swift} \citep{gehrels04} is currently a major player in the search for the electromagnetic counterpart of a neutrino event. In addition, the possibility to combine simultaneous information in the optical-UV bands collected by the UVOT telescope with the observations collected by the XRT instrument is an added value in the search for counterparts. Rapid follow-up observations of neutrino-emitting AGN can also be performed with the Space-based multi-band astronomical Variable Objects Monitor \citep[SVOM;][]{atteia22}. Another potential future X-ray telescope important for studying the AGN-neutrino connection may be the Advanced X-ray Imaging Satellite \citep[AXIS;][]{koss25}. The enhanced sensitivity of AXIS can enable continuous monitoring of AGN with short exposure times, providing a detailed understanding of their baseline X-ray emission. Moreover, the rapid response time of AXIS ($<$ 2 hours, comparable to \textit{Swift}) is ideal for high-cadence  follow-up observations of candidate sources in the direction of neutrino alerts, allowing timely identification and spectral characterization of potential neutrino counterparts. Future observations with the \textit{NewAthena} mission \citep{cruise25} will allow us to obtain a detailed characterization of the neutrino-emitting AGN in X-rays. Finally, the Imaging X-ray Polarimetry Explorer \citep[IXPE;][]{weisskopf22} can provide unique data on the geometry of magnetic fields and particle acceleration mechanisms in blazars. X-ray polarization is used as a tool to distinguish between purely leptonic and hadronic (neutrino-producing) emission models. High, stable polarization often suggests acceleration in ordered shocks, while rapid variations indicate magnetic turbulence. IXPE observations of blazars seem to support models of shock acceleration in stratified jets, identifying the presence of helical magnetic fields that could facilitate neutrino production \citep[e.g.,][]{maksym25}. IXPE observations are also important for characterizing the geometry of the hot X-ray corona  of Seyfert galaxies \citep[e.g.,][]{gianolli23}. 

In the UV band, a future facility that can be important in operating in synergy with neutrino telescopes will be the Ultraviolet Explorer \citep[UVEX;][]{kulkarni21}, an upcoming wide-field ultraviolet space telescope from NASA scheduled to launch in 2030. UVEX with respect to other telescopes will have a combination of a wide field of view, sensitivity, and rapid response time needed to perform follow-up observations of neutrino-emitting AGN. 

Optical follow-up of flaring AGN has already demonstrated to be successful in characterizing their activities, in particular in a broader multi-frequency context. However, the situation is more complex in the case of the investigation of the neutrino-AGN connection. Important properties for discovering optical counterparts of high-energy neutrinos are: a large field of view, high sensitivity to detect also fainter objects, deep high-cadence survey baseline to characterize both historical and recent activity of the candidate AGN, multi-band color information for a better identification of the counterpart via color, color evolution, and rapid light curve evolution. Currently operating neutrino follow-up programs include ASAS-SN \citep{necker22}, DECam \citep{flaugher15}, Kiso and Subaru \citep{morokuma24}, Pan-STARRS1 \citep{kankare19}, and ZTF \citep{stein23}. However, none of these projects meets all the properties reported above. In this context, follow-up with the Vera C. Rubin Observatory \citep{ivezic19} can be transformational, providing a more comprehensive view of the multi-messenger sky with optical. 
Currently, the main issue of neutrino follow-up with Vera Rubin is related to the fact that most of the neutrino alerts come from IceCube in the Northern Sky, only partially covered by the Legacy Survey of Space and Time (LSST). Vera Rubin will be even more important when new neutrino facilities become completely operational in the Southern hemisphere, making it more accessible to Vera Rubin. Multicolor observations of candidate neutrino emitting AGN can also be obtained with space-based missions such as Euclid \citep{mellier25}, Spectro-Photometer for the History of the Universe, Epoch of Reionization and Ices Explorer (SPHEREx) \citep{crill20}, and the future Nancy Grace Roman telescope \citep{troxel23}.

Radio monitoring of candidate neutrino-emitting AGN has been performed from 5 to 37 GHz in recent years with single-dish telescopes such as OVRO, Mets\"ahovi \citep{hovatta21}, and Effelsberg \citep{eppel24}. Observations at mm and sub-mm with the Atacama Large Millimeter Array \citep[ALMA;][]{wooten09} can provide high-resolution images of the gas and dust surrounding the SMBH at the centers of the AGN. In particular, ALMA observations can be important in understanding whether AGN jets interact with nearby gas or dust, an interaction that could significantly increase the flux of neutrinos produced. 
Future facilities such as ngVLA \citep{selina18} and ngEHT \citep{doeleman23} will provide detailed images of the jet and core regions of these candidates, down to (sub)parsec-scale resolutions, allowing us to study the mechanisms responsible for high-energy photon and neutrino events and investigate structural changes connected to neutrino emission \citep[see e.g.,][]{lico23,kovalev23}. Finally, the Square Kilometre Array Observatory (SKAO) \citep{braun15} can provide important information for different aspects of the neutrino-AGN connection: i) SKAO will allow us to determine common radio features among neutrino sources, helping to discriminate between various emission mechanisms; ii) thanks to its ability to detect extremely faint AGN (both radio-loud and radio-quiet), SKAO will help map AGN activity over cosmic time; iii) high spatial resolution observations will allow us to study the internal regions of accretion discs and the ejection processes (winds and jets) where the origin of neutrinos is hypothesized.
 
\section{Concluding remarks}\label{sec8}

The detection of an astrophysical flux of neutrinos in the TeV--PeV band has opened up a new exciting window to learn more about extreme cosmic accelerators and neutrino interactions at the highest energies \citep{aartsen13, baikal23}. The identification of the sites of cosmic-ray acceleration in the Universe is one of the most intriguing  questions in astrophysics \citep[e.g.,][]{blandford14}. A part of the neutrino flux arises from our own Milky Way, but most neutrinos have an extragalactic source. In particular, the isotropy of the neutrino arrival directions strongly favours an extragalactic origin for these neutrinos, with the AGN, both jetted and non-jetted ones, being a promising class of neutrino-emitting sources. However, the challenge in identifying sources is caused by low signal statistics along with relatively large uncertainties in the angular localization. In this context, simultaneous electromagnetic observations are essential for identifying neutrino sources.

The coincidence of the arrival of the high-energy neutrino IceCube-170922A with the spatial position and flaring activity of the blazar TXS\,0506+056 was the first compelling evidence that an AGN can be a neutrino-emitting source. However, blazars have been investigated as neutrino-emitting sources using a stacking analysis but significant correlations have not been obtained. Further correlation searches between high-energy neutrinos and known extragalactic sources have been performed, with only two sources for which a statistically significant correlation has been established so far: the blazar TXS\,0506+056 and the Seyfert II galaxy NGC\,1068. 
The search for correlation between neutrino arrival and known sources has been performed by analyzing single high-energy neutrino emitted at a time or by stacking sources belonging to the same class/sub-class of objects. In the former case, individual sources are tentatively associated to high-energy neutrinos, but statistical trials have to be considered for the look-elsewhere effect, and this can reduce the statistical significance of the detection if multiple sources are considered. In the latter case, a stacking method has been used, but a different weighting of the fluxes can be assumed in the search, making the analysis sensitive to this assumption. Moreover, time-integrated and time-dependent searches have been performed. Time-integrated searches are performed in a background-dominated regime; thus enough data have to be accumulated before a signal emerges. Time-dependent searches of individual neutrino events to be associated with extragalactic sources need that the neutrino events have an energy sufficiently high as the atmospheric background starts to be subdominant (e.g., $E > 100\, \mathrm{TeV}$). In both cases, the number of high-energy events accumulated and the speed with which the detector can accumulate events are fundamental.

The neutrino production requires the acceleration of protons and the presence of target photons. These two requirements are satisfied close to the black hole at the center of the AGN, as indicated by the detection of neutrinos by NGC\,1068. Neutrinos can be produced by the X-ray corona but also regions of high hydrogen density near the core and UV photons associated with the accretion disc. However, the corona is fundamental to suppress GeV--TeV emission by absorbing the pionic $\gamma$-rays. In this context, NGC\,1068 can be defined as the archetypal hidden neutrino AGN. Future observations in the MeV energy range of NGC\,1068 and other bright Seyfert can be important in determining the coronal activity, where the absorbed GeV--TeV photons should be reprocessed. It is worth mentioning also that both NGC\,1068 and TDE associated with neutrino so far are near-Eddington accretion systems, and hidden neutrino sources, suggesting that in those classes of sources the neutrino emission can be produced by the same mechanisms \citep[e.g.][]{murase20}. In case of blazars, neutrinos seems to be primarily produced within the relativistic jet or very close to its origin, specifically in the central sub-pc regions near the SMBH.

The contribution of individual sources to the diffuse high-energy neutrino flux remains minimal, representing $\sim$ 1\% of the total flux \citep[e.g.][]{abbasi22}. Furthermore, the contribution is not considered significant even when specific classes of AGN are accounted for, remaining below 15\%. In contrast, the Galactic plane contributes $\sim$10\% at 30 TeV \citep{abbasi23c}. However, recent studies suggest that AGN may account for a significant portion, or even the majority, of the observed flux in specific energy ranges. In particular, non-jetted AGN can theoretically contribute up to 95\% of the IceCube Observatory diffuse neutrino flux in the 1--100 TeV energy range, while jetted AGN (blazars) are expected to provide a significant contribution at $E > 1\, \mathrm{PeV}$ \citep[e.g.,][]{padovani24}.


Currently, neutrino facilities are relatively limited in terms of angular resolution and sensitivity, thus it is hard to have an unambiguous association between neutrino emission and a single source, and information over the entire electromagnetic spectrum about the emission at the time of the neutrino arrival but also long-term monitoring of a sample of candidate emitting AGN can provide further information. Statistical search for neutrino emitters is now possible thanks to the data accumulated by IceCube. With the next-generation neutrino experiments we will push sensitivity, angular resolution, and energy band, dramatically expanding the wealth of information available. In addition, due to the different locations and capabilities, a joint effort of different neutrino telescopes will improve searches for extragalactic neutrino-emitting source. 

It will also be important for a rapid coordination of multi-messenger and multi-frequency facilities that neutrino facilities implement an automated release of fully-machine-readable neutrino alerts that include localization and related systematic uncertainties, a simple data quality flag in order to trigger automatic decision algorithm for activating follow-up observations. Another aspect that would be important for making a further step in our knowledge of the physics of neutrino-emitting AGN is an improvement of numerical simulations on source dynamics and particle-acceleration processes with a direct connection to the observational evidence that we are acquiring for a coherent picture of the multi-messenger scenarios related to the neutrino physics of the AGN. 

Multi-messenger astrophysics is expected to reach full maturity in the second half of the next decade, with new neutrino facilities, the deployment of the third generation of ground-based gravitational-wave detectors such as the Einstein Telescope \citep{abac26}, and the space-borne gravitational-wave observatory LISA \citep{colpi24}. Only a unified community approach will allow substantial synergy, enabling efficient characterization through multi-wavelength analysis. Therefore, in the future, it will be fundamental to integrate the neutrino telescopes with the multi-messenger astronomy community to have strictly simultaneous neutrino, electromagnetic, and, if possible, gravitational-wave observations of the significant events detected by the current and future neutrino observatories to improve the sensitivity of joint emitters of these three messengers and discover more correlations in near-real time. Thanks to forthcoming neutrino observatories such as KM3Net, Baikal-GVD, IceCube-Gen2 operating in parallel with a new generation of sensitive all-sky X-ray, MeV, and GeV--TeV satellites, and GW facilities, neutrino astrophysics can make a significant step forward in our knowledge of the role of the different class of sources, and AGN in particular, in neutrino and UHECR emission and the interconnection between the different messengers in the Universe. 



\bmhead{Acknowledgements}
I would like to thank the referees for the careful reading of the manuscript and the detailed comments and suggestions that helped to improve the review.

\section*{Declarations}

\bmhead{Conflict of interest}
The authors declare no conflict of interest.

\phantomsection
\addcontentsline{toc}{section}{References}
\bibliography{Neutrino_DAmmando.bib}

\end{document}